\pdfoutput=1

\documentclass[11pt]{article}

\usepackage{latex/acl}

\usepackage{times}
\usepackage{svg}
\usepackage{natbib}
\usepackage{seqsplit}
\usepackage{graphicx}
\usepackage{latexsym}
\usepackage{booktabs}
\usepackage{pifont}
\usepackage{multirow}
\usepackage{array}

\usepackage[T1]{fontenc}

\usepackage[utf8]{inputenc}

\usepackage{microtype}
\usepackage{amsmath}
\usepackage{mathtools}
\usepackage{makecell}

\usepackage{inconsolata}
\usepackage{amssymb}
\usepackage{xspace,mfirstuc,tabulary}

\newcommand{\datasetname}{DebugBench\xspace}

\newcommand{\codefont}[1]
{{\fontfamily{qcr}\selectfont #1}}

\usepackage{geometry}
\usepackage{xcolor}
\usepackage{tikz}
\usepackage{enumitem}
\usepackage{lipsum}
\usepackage[most]{tcolorbox}

\definecolor{trc_lightgrey}{RGB}{240, 240, 240}
\definecolor{trc_darkgrey}{RGB}{169, 169, 169}
\definecolor{darkerGrey}{RGB}{105, 105, 105}
\definecolor{trc_lightgreen}{RGB}{245, 250, 241}
\definecolor{trc_darkgreen}{RGB}{207, 228, 186}
\definecolor{trc_lightblue}{RGB}{241, 245, 250}
\definecolor{trc_darkblue}{RGB}{186, 198, 230}
\definecolor{trc_lightred}{RGB}{251, 241, 241}
\definecolor{trc_darkred}{RGB}{236, 185, 191}

\usepackage{listings}
\usepackage{xcolor}

\newtcolorbox{prompt}{
  colback=trc_lightgrey, 
  colframe=trc_darkgrey, 
  colbacktitle=darkerGrey, 
  enhanced,
  boxrule=0pt,
  after skip=0cm,
  before skip=0.3cm,
  right skip=0cm,
  breakable,
  fonttitle=\small\bfseries, 
  fontupper=\small\linespread{1.25}\selectfont, 
  toprule=0pt,
  bottomrule=0pt,
  rightrule=0pt,
  leftrule=4pt,
  arc=0mm,
  skin=enhancedlast jigsaw,
  sharp corners,
  boxed title style={
    frame code={ 
    }
  }
}

\newtcolorbox{prompt_blue}{
  colback=trc_lightblue, 
  colframe=trc_darkblue, 
  colbacktitle=darkerGrey, 
  enhanced,
  boxrule=0pt,
  after skip=0cm,
  before skip=0.3cm,
  right skip=0cm,
  breakable,
  fonttitle=\small\bfseries, 
  fontupper=\small\linespread{1.25}\selectfont, 
  toprule=0pt,
  bottomrule=0pt,
  rightrule=0pt,
  leftrule=4pt,
  arc=0mm,
  skin=enhancedlast jigsaw,
  sharp corners,
  boxed title style={
    frame code={ 
    }
  }
}

\newtcolorbox{prompt_red}{
  colback=trc_lightred, 
  colframe=trc_darkred, 
  colbacktitle=darkerGrey, 
  enhanced,
  boxrule=0pt,
  after skip=0cm,
  before skip=0.3cm,
  right skip=0cm,
  breakable,
  fonttitle=\small\bfseries, 
  fontupper=\small\linespread{1.25}\selectfont, 
  toprule=0pt,
  bottomrule=0pt,
  rightrule=0pt,
  leftrule=4pt,
  arc=0mm,
  skin=enhancedlast jigsaw,
  sharp corners,
  boxed title style={
    frame code={ 
    }
  }
}

\newtcolorbox{prompt_green}{
  colback=trc_lightgreen, 
  colframe=trc_darkgreen, 
  colbacktitle=darkerGrey, 
  enhanced,
  boxrule=0pt,
  after skip=0cm,
  before skip=0.3cm,
  right skip=0cm,
  breakable,
  fonttitle=\small\bfseries, 
  fontupper=\small\linespread{1.25}\selectfont, 
  toprule=0pt,
  bottomrule=0pt,
  rightrule=0pt,
  leftrule=4pt,
  arc=0mm,
  skin=enhancedlast jigsaw,
  sharp corners,
  boxed title style={
    frame code={ 
    }
  }
}

\title{\raisebox{-1.0ex}{\includegraphics[width=0.8cm]{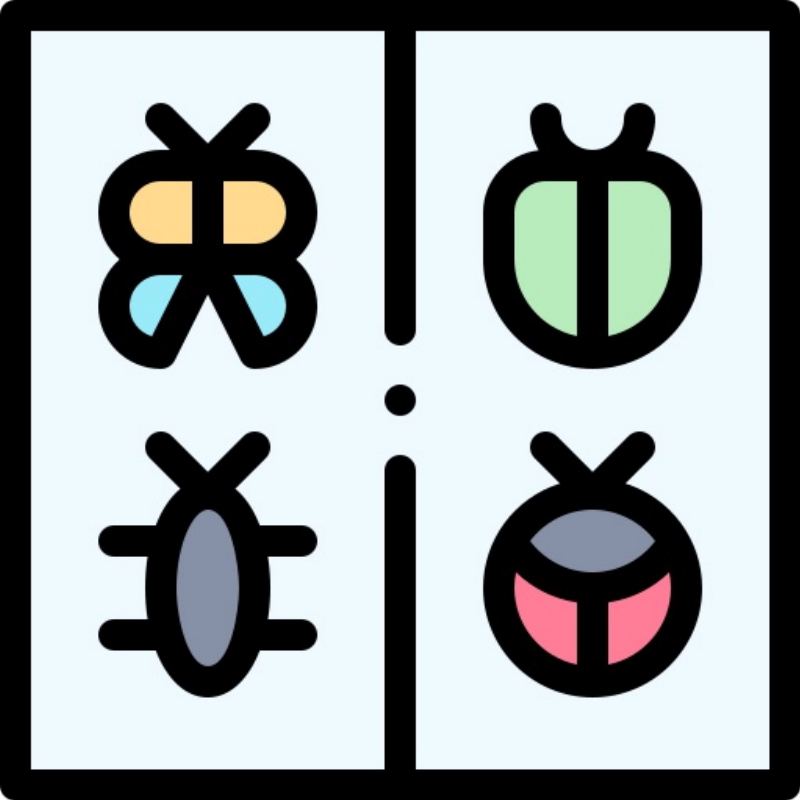}}\hspace{0.1cm}
DebugBench: \\ \vspace{0.3cm} Evaluating Debugging Capability of Large Language Models}

\usepackage{authblk}

\author{
\textbf{Runchu Tian}$^{1}$\thanks{\ \ indicates equal contribution.}\hspace{0.5cm}
\textbf{Yining Ye}$^{1*}$\hspace{0.5cm}
\textbf{Yujia Qin}$^{1}$\hspace{0.5cm}
\textbf{Xin Cong}$^{1}$\hspace{0.5cm}
\textbf{Yankai Lin}$^{2}$\thanks{\hspace{0.15cm}Corresponding Author}\hspace{0.5cm}
\\
\textbf{Yinxu Pan}$^{3}$\hspace{0.5cm}
\textbf{Yesai Wu}$^{3}$\hspace{0.5cm}
\textbf{Haotian Hui}$^{4}$\hspace{0.5cm}
\textbf{Weichuan Liu}$^{4}$\hspace{0.5cm}
\textbf{Zhiyuan Liu}$^{1\dag}$\hspace{0.5cm}
\textbf{Maosong Sun}$^{1\dag}$
\\
$^1$Tsinghua University\hspace{0.5cm} $^2$Renmin University of China\hspace{0.5cm}
$^3$ModelBest Inc.
$^4$Siemens AG.
\\
\texttt{\small trc20@mails.tsinghua.edu.cn, yeyn2001@gmail.com}

}

\begin{document}
\maketitle

\begin{abstract}

Large Language Models (LLMs) have demonstrated exceptional coding capability. However, as another critical component of programming proficiency, the debugging capability of LLMs remains relatively unexplored. Previous evaluations of LLMs' debugging ability are significantly limited by the risk of data leakage, the scale of the dataset, and the variety of tested bugs. To overcome these deficiencies, we introduce `DebugBench', an LLM debugging benchmark consisting of 4,253 instances. It covers four major bug categories and 18 minor types in C++, Java, and Python. To construct DebugBench, we collect code snippets from the LeetCode community, implant bugs into source data with GPT-4, and assure rigorous quality checks. We evaluate two commercial and four open-source models in a zero-shot scenario. We find that (1) while closed-source models exhibit inferior debugging performance compared to humans, open-source models relatively lower pass rate scores; (2) the complexity of debugging notably fluctuates depending on the bug category; (3) incorporating runtime feedback has a clear impact on debugging performance which is not always helpful. As an extension, we also compare LLM debugging and code generation, revealing a strong correlation between them for closed-source models. These findings will benefit the development of LLMs in debugging. Our code and dataset are open-sourced via \href{https://github.com/thunlp/DebugBench}{GitHub repository} and \href{https://huggingface.co/datasets/Rtian/DebugBench}{Hugging Face dataset}.

\end{abstract}

\section{Introduction}
\label{sec:Introduction}

Large language models (LLMs) have demonstrated exceptional code generation abilities. LLM-based coding methods~\citep{zhou2023language, shinn2023reflexion} have achieved human-level performance on benchmarks like HumanEval~\citep{chen2021evaluating} and MBPP~\citep{austin2021program}. LLMs have also become the core engine of practical programming assistance applications like \citet{github_copilot}.
\begin{figure}[!t]
  \centering
  \includegraphics[width=0.48\textwidth]
  {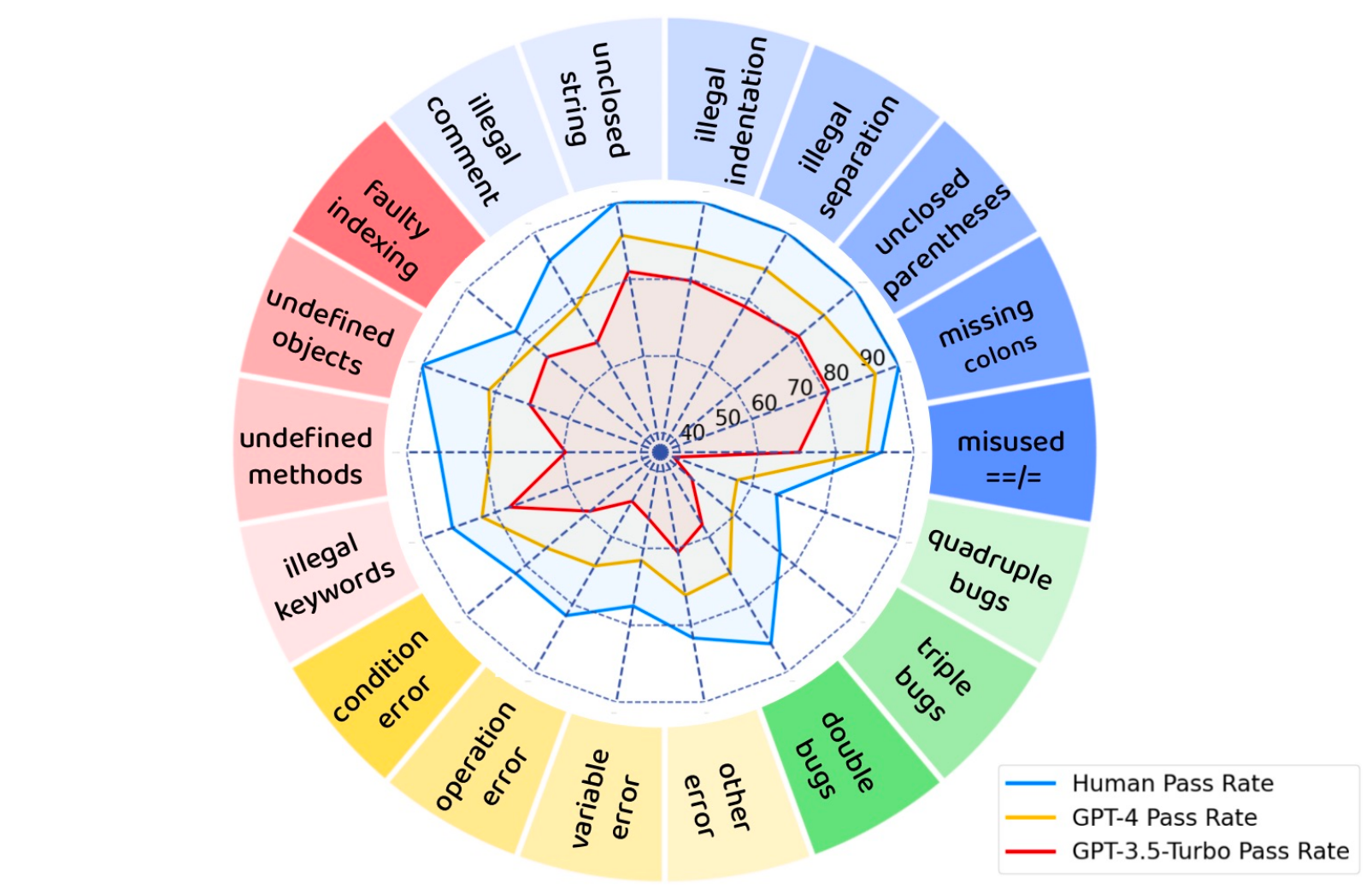}
  \caption{This figure illustrates the comparative debugging performance of \codefont{gpt-3.5-turbo-0613}~\citep{openai2022chatgpt}, \codefont{gpt-4-0613}~\citep{openai2023gpt4} and human proficiency across various bug categories. Evaluations are also performed on open-sourced models which are not exhibited in this figure.}
  \label{fig:radar}
\end{figure}
Similar to code generation, debugging\footnote{We use the colloquial terms `bug', `buggy', and `debug' to refer to programming errors that cause unintended runtime behavior, to code containing these errors, and to the process of locating and eliminating them, respectively.} is also a crucial component in programming, consuming 35-50\% of the development duration and 50-75\% of the total budget~\citep{mcconnell2004code}. However, unlike coding, the debugging abilities of LLMs remain relatively unexplored. 

\begin{table*}[!ht]
\centering
\scalebox{0.8}{
\begin{tabular}{lrcccc}
\toprule[1.1pt]
\textbf{Work} & \textbf{Test Scale} & \textbf{Against Data Leakage} & \textbf{Bug Type Diversity} & \textbf{Model Diversity} & \textbf{Scenario Diversity}\\
\midrule[0.7pt]
\citet{prenner2022can} & 40 & \textcolor{red}{\ding{55}} & \textcolor{red}{\ding{55}} & \textcolor{red}{\ding{55}} & \textcolor{red}{\ding{55}} \\
\citet{sobania2023analysis} & 40 & \textcolor{red}{\ding{55}} & \textcolor{red}{\ding{55}} & \textcolor{red}{\ding{55}} & \textcolor{red}{\ding{55}} \\
\citet{xia2023conversational} & 60 & \textcolor{red}{\ding{55}} & \textcolor{red}{\ding{55}} & \textcolor{green}{\ding{51}} & \textcolor{green}{\ding{51}}\\
\citet{zhang2023critical} & 151 & \textcolor{green}{\ding{51}} & \textcolor{red}{\ding{55}} & \textcolor{green}{\ding{51}} & \textcolor{green}{\ding{51}} \\
\midrule
DebugBench  & 4,253 & \textcolor{green}{\ding{51}} & \textcolor{green}{\ding{51}} & \textcolor{green}{\ding{51}} & \textcolor{green}{\ding{51}} \\
\bottomrule[1.1pt]
\end{tabular}}
\caption{Limitations of prior studies in LLM debugging. We introduce DebugBench, a new LLM debugging benchmark to overcome these deficiencies. }
\label{tab:evaluation_comparison}
\end{table*}

One primary obstacle in code debugging research is the lack of evaluation benchmarks. While some basic evaluations~\citep{prenner2022can, sobania2023analysis, xia2023conversation, zhang2023critical} verify the effectiveness of LLM-based debugging methods, these evaluations have notable limitations that prevent us from comprehensively assessing the debugging capabilities of LLMs as exhibited in Table \ref{tab:evaluation_comparison}. 
First, as \citet{zhang2023critical} revealed, existing debugging benchmarks~\citep{just2014defects4j, lin2017quixbugs} have been more or less leaked to the pre-training data of popular LLMs via web scraping and other means. For instance, ChatGPT~\citep{chatgpt2023} can enumerate all the projects in Defects4J~\cite{just2014defects4j}. While it's challenging to ascertain the exposure due to a lack of training details, there's a significant risk of \textbf{data leakage}.
Second, all existing debugging evaluations have been limited to a very \textbf{small scale}, ranging from 40 to 151 examples, which may hurt the generalizability of the assessments. Third, existing works reported a general pass rate across various bug categories instead of \textbf{differentiating various bug types}. Analyzing the variations in performance across different bug types can reveal the bottlenecks and guide focused improvements in LLM debugging.

To overcome these deficiencies, we create DebugBench, a dataset of 4,253 instances for LLM debugging evaluation. We first collect code solution snippets from \citet{leetcode2023}, a popular programming challenge platform.  To reduce the risk of data leakage, we ensure all of the instances in DebugBench are released after July 2022, which is beyond the pre-training data cutoff date of tested models. For fine-grained evaluation of various bug types, we develop a bug taxonomy based on \citet{barr2004find}'s classification criteria. The classification encompasses four major bug categories: Syntax, Reference, Logic, and Multiples, along with 18 minor types as illustrated in Figure \ref{fig:radar}. Subsequently, we prompt GPT-4~\citep{openai2023gpt4} to implant bugs into the code solutions in pursuit of sufficient data scales for each bug type. We cover snippet-level code in C++, Java, and Python. To ensure integrity, we conduct automatic filtering and manual inspection.

As shown in Figure \ref{fig:radar}, we evaluate two closed-source language models, \codefont{gpt-4-0613}~\citep{openai2022chatgpt} and \codefont{gpt-3.5-turbo-0613}~\citep{openai2023gpt4}, along with four open-source models: \codefont{CodeLlama-7b-Instruct}~\citep{rozière2023code}, \codefont{Llama-3-8B-Instruct}~\citep{llama3_website}, \codefont{DeepSeek-Coder-33B-Instruct}~\citep{deepseek-coder} and \codefont{Mixtral-8x7B-Instruct}~\citep{jiang2024mixtral} in zero-shot scenarios. Our empirical study reveals: (1) \textbf{LLM debugging falls short of human performance.} Open-source models attain low pass rate scores for debugging queries. Closed-source LLMs surpass open-source ones but still fall short of human-level performance; (2) \textbf{The difficulty of fixing different types of errors differs.} Multiple errors and logical errors are significantly more challenging to repair than syntax and reference errors; (3) \textbf{Runtime feedback has a clear impact on LLM's debugging performance but is not always helpful.} While runtime feedback consistently boosts the debugging performance of syntax and reference bugs, the feedback information is unhelpful for logic errors.

To gain deeper insights into the overall programming capabilities of LLMs, we also compare closed-source models' performance on debugging and code generation. Experimental results indicate that for closed-source models: (1) fixing syntax or reference errors is generally easier than code generation, while repairing logical or multiple errors can be equally hard or even harder; (2) the debugging and code generation performance of LLMs are correlated, which indicates the abilities of LLMs to approach these two tasks are positively related.
All these findings are crucial for comprehending the debugging capabilities of LLMs and developing more comprehensive code models.

\begin{figure*}
  \centering
  \includegraphics[width=\textwidth]{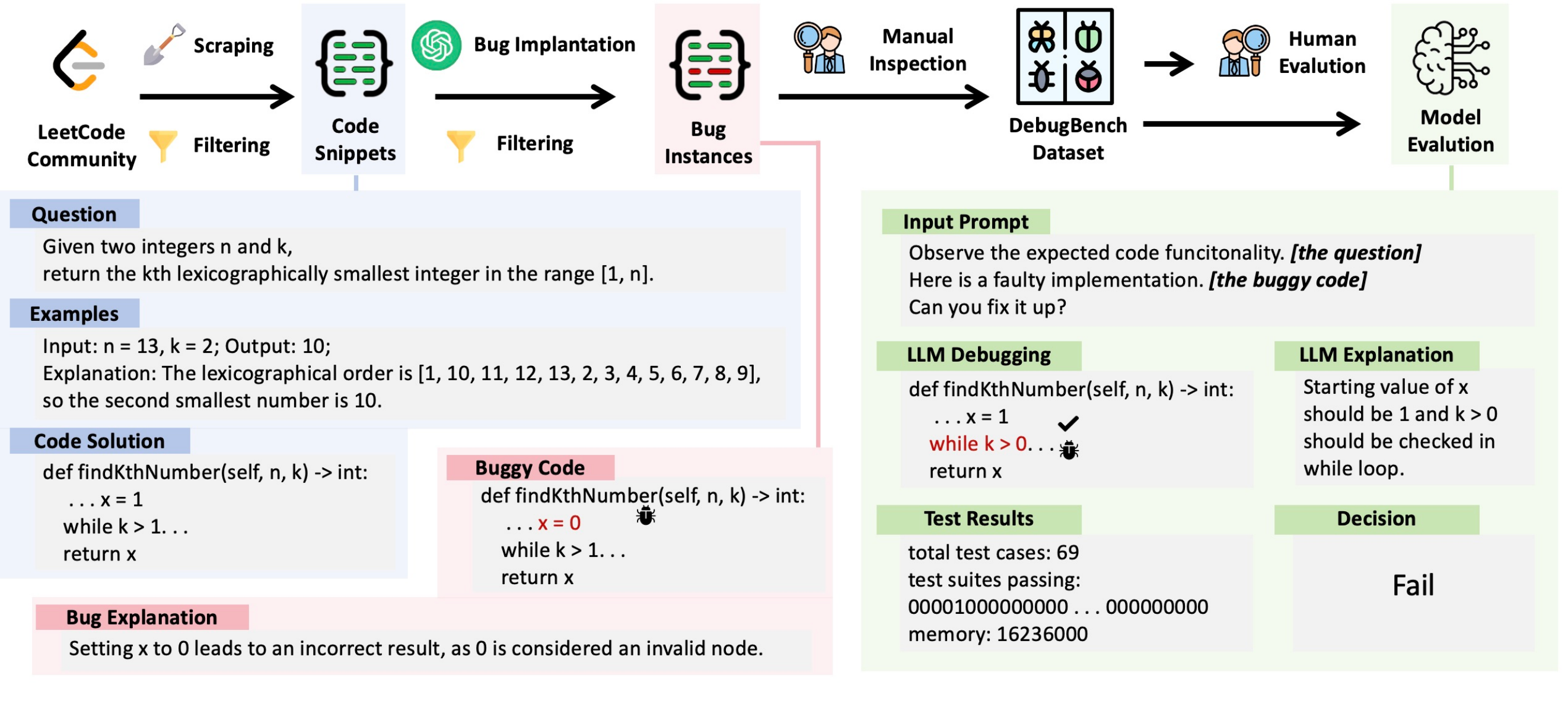}
  \caption{This figure illustrates the construction of DebugBench. We first collect code snippets from \citet{leetcode2023} community, then employ GPT-4~\citep{openai2023gpt4} for bug implantation and finally conduct human / LLM evaluation on the benchmark. Automatic filtering and final human inspection are conducted to ensure integrity of the benchmark. The figure also provides qualitative cases for code snippets, bug instances, and evaluation samples. More examples are accessible in Appendix \ref{sec:Examples of Synthesized Bugs}.}
  \label{fig:construction}
\end{figure*}

\section{Benchmark Construction}

As illustrated in Figure \ref{fig:construction}, to construct DebugBench, we first collect questions, code snippets, and examples from \citet{leetcode2023} community, then employ GPT-4~\citep{openai2023gpt4} for bug implantation. To ensure the integrity of the benchmark, we conduct automatic filtering and final human inspection.

\subsection{Formulation of Debugging}
Consider the input-output pairs $(x_i, y_i)$ where each $x_i$ is a program input and $y_i$ is the corresponding desired output, together they compose a set $R$ that defines the programming problem.

Let $a_\theta(x)=y$ denote a program $a$, based on a code script $\theta$, that maps an input $x$ to an output $y$. We identify a code script $\theta$ that exists bugs if there exists a pair $(x_i, y_i) \in R$ such that $a_{\theta}(x_i) \neq y_i$.

Consequently, an ideal debugger $D$ that rectifies any buggy code from $\theta$ to $\theta^*$ should satisfy that $ D(\theta) = \theta^* \ s.t. \ \forall (x_i, y_i) \in R,\  a_{\theta^*}(x_i) = y_i$. Debugging can be regarded as the converting process of debugger $D$.

\subsection{Source Data Collection}
We collect 3,206 samples from user-submitted solutions to specific programming challenges on~\citet{leetcode2023}. Each sample contains the question, solution code, examples, and release date. We utilize GPT-2~\citep{radford2019language} tokenizer to tokenize these instances and report an average token length of 468.1 tokens, a typical length scale of code snippets. All of instances were released after June 2022, with an average release date of April 2023. This minimizes the risk of data leakage\footnote{The cutoff date of pretraining data for \codefont{gpt-3.5-turbo-0613} and \codefont{gpt-4-0613} is officially  September 2021~\citep{chatgpt2023}.}~\citep{zhang2023critical}.

Apart from reducing data leakage, our choice of LeetCode is driven by two other reasons: (1) LeetCode offers sufficiently challenging code problems even for state-of-the-art LLMs like GPT-4 \citep{shinn2023reflexion}; (2) LeetCode provides comprehensive test suites that facilitate automated evaluation, while other data sources like \citet{github} may suffer from arduous human labor~\citep{hu2023instructcoder} or incomplete test suites. A qualitative example of scraped code snippets can be found in Figure~\ref{fig:construction}.

We select the three most popular programming languages \citep{tiobe-index}, C++, Java, and Python3, to reflect the LLM debug capability in real-world scenarios. Our dataset comprises 1,438 instances in C++, 1,401 in Java, and 1,414 in Python.

\subsection{Bug Implantation}

After collecting source data from \citet{leetcode2023}, we adopt GPT-4 \citep{openai2023gpt4} to implant bugs into code snippets. For implanting single errors (one bug in one code snippet), we prompt GPT-4 \citep{openai2023gpt4} with the correct code, desired bug type and instruct the model to generate a buggy version of the input code and a few sentences of explanation on the inserted bug. To implant multiple errors, we adopt rule-based merging based on single errors, which is similar to the merge operation in version control systems. The prompt we use for bug implantation can be found in Appendix \ref{sec:prompts}.

We instruct GPT-4 to add diverse types of bugs into code snippets. Based on the bug classification criteria from \citet{barr2004find}, we categorize the bug into 4 major categories and 18 minor types. Table \ref{tab:type_numbers} depicts the scope of bug types in \datasetname. This diversity enables a thorough investigation of LLMs' ability to debug a wide array of programming errors. The definition of each minor type can be found in Appendix \ref{sec:Definition and Distribution of Each Type of Bugs}. 

We choose bug synthesis rather than bugs from traditional Debugging datasets like Defects4J \citep{just2014defects4j} and QuixBugs \citep{lin2017quixbugs} in pursuit of a vast degree of freedom in error diversity design and lower risk of data leakage.
The feature schema of generated instances be found in Appendix \ref{sec:feature schema}.

\begin{table}[!t]
\centering
\small 
\begin{tabular}{ccr}
\toprule[1.1pt]
\textbf{Type} & \textbf{Minor Type} &\textbf{Number}\\
\midrule[0.7pt]
\multirow{7}{*}{Syntax} & misused ==/= & 137\\
& missing colons & 129\\
& unclosed parentheses & 133\\
& illegal separation & 68\\
& illegal indentation & 45\\
& unclosed string & 125\\
& illegal comment & 124\\
\midrule[0.7pt]
\multirow{4}{*}{Reference}& faulty indexing & 206 \\
& undefined objects & 187\\
& undefined methods & 167\\
& illegal keywords & 124\\
\midrule[0.7pt]
\multirow{4}{*}{Logic}& condition error & 260 \\
&operation error & 180\\
&variable error & 100\\
&other error & 50\\
\midrule[0.7pt]
\multirow{3}{*}{Multiple}& double bugs & 750 \\
&triple bugs & 750\\
&quadruple bugs & 718\\
\bottomrule[1.1pt]
\end{tabular}
\caption{Bug types and their distribution in DebugBench.}
\label{tab:type_numbers}
\end{table}


\subsection{Quality Control}
\label{sec:Quality Control}
To ensure the quality of \datasetname, we conduct automatic filtering and manual inspection. 

\paragraph{Automatic Filtering.}
First, we filter the source data collected from \citet{leetcode2023}. We design the following automatic filtering criteria: (1) The code solution must be correct, that is, to pass the whole corresponding test suites. (2) The instances must contain necessary information like programming language, release time, and question id. (3) The release date of code snippets must be no earlier than July 2022, the official knowledge cutoff date of two closed-source models~\citep{chatgpt2023} in case of data leakage. 72.1\% of the user-submitted code snippets pass this automatic filtering. Second, we filter the data synthesized by GPT-4~\citep{openai2023gpt4} since the LLM occasionally fails to perform bug implantation as expected. We again establish automatic filter criteria: (1) The code with implanted bugs must fail certain test cases to confirm its erroneous nature. (2) The buggy code should not include in-line comments that could leak information about the bug. (3) The explanation for the bug must be thorough and relevant to the assigned bug type. Following these criteria, 79.2\% of the 3,000 bug-implanted instances pass the filtering process. 

\paragraph{Manual Inspection.}
After automatic generation and filtering, we manually inspect the quality of DebugBench. We apply three criteria for manual inspection: (1) \textbf{Bug Validity}: The bugs must cause the intended malfunction, fail specific test cases, and align with the assigned bug type and description. (2) \textbf{Sensitive Information Security}: The instances must be devoid of sensitive data, such as personal information. (3) \textbf{Scenario Alignment}: The bugs should resemble those found in actual code debugging scenarios and should not include obvious clues, like comments indicating the bug's location.

We hire three programmers with over four years of experience in programming to conduct the manual inspection on 180 cases over two hours each after training on 30 cases. Their review reveals that the DebugBench benchmark is of high quality as exhibited in Table \ref{tab:manual_inspection}. Failing cases can be found in Appendix \ref{sec:Cases Failing Manual Inspection.}.

\begin{table}[!t]
\centering
\small
\begin{tabular}{cr}
\toprule[1.1pt]
\textbf{Criteria} & \textbf{Pass Rate/\%}\\
\midrule[0.7pt]
Bug Validity & 97.4\\
Sensitive Information Security & 100.0\\
Scenario Alignment & 93.2\\
All Three criteria & 92.1\\
\bottomrule[1.1pt]
\end{tabular}
\caption{Results of manual inspection of DebugBench.}
\label{tab:manual_inspection}
\end{table}

\subsection{Analysis of Bug Realism}
To ensure the diversity of bug types and mitigate the risk of data leakage, we adopted a method of collecting code snippets from programming communities and subsequently injecting bugs into them to construct our test data. However, this approach has raised concerns regarding the realism of synthetically generated bugs compared to those encountered in real-world scenarios. This discrepancy may result in DebugBench evaluations not adequately reflecting a model’s capability to fix bugs in actual situations.

\begin{table}[!t]
\centering
\small
\scalebox{0.9}{
\begin{tabular}{ll}
\toprule[1.1pt]
\textbf{Bugs from QuixBugs} & \textbf{Bugs from DebugBench}\\
\midrule[0.7pt]
Incorrect assignment operator           & misused ==/!=       \\
Incorrect variable                      & variable error            \\
Incorrect comparison operator           & misused ==/!=       \\
Missing condition                       & condition error           \\
Missing/added +1                        & operation error           \\
Variable swap                           & variable error            \\
Incorrect array slice                   & faulty indexing           \\
Variable prepend                        & variable error            \\
Incorrect method called                 & undefined methods   \\
Incorrect field dereference             & undefined objects         \\
Missing arithmetic expression           & operation error           \\
Incorrect data structure constant       & None    \\
Missing function call                   & None                      \\
Missing line                            & None                      \\
\bottomrule[1.1pt]
\end{tabular}
}
\caption{The correspondence of bug type of QuixBugs~\citep{lin2017quixbugs} in DebugBench. The bug types in QuixBugs are largely encompassed by those in DebugBench.}
\label{tab:class_compare}
\end{table}

To address these concerns, we conducted a comparative analysis between the characteristics of bugs in DebugBench and those from other real-world bug datasets. Specifically, we examined QuixBugs~\citep{lin2017quixbugs}, which contains bugs from the Quixey Challenge based on real-world scenarios. Our findings are as follows: (1) \textbf{Bug Types}: DebugBench, which significantly greater diversity in bug types, includes the majority of bug types present in QuixBugs as detailed in Table~\ref{tab:class_compare}. (2) \textbf{Bug Recognition}: We utilized GPT-4~\citep{openai2023gpt4} and Claude 3 Sonnet~\citep{anthropic2024claude} to identify bugs of the same type in both datasets and observed the recognition success rate, as presented in Table~\ref{tab:bug_comparison}. We conducted a total of 420 comparisons for each judger. Details about this analysis can be found in Appendix~\ref{sec: Comparison of Bug Realism}.

\begin{table}[htbp]
\centering
\small
\scalebox{0.9}{
\begin{tabular}{lccc}
\toprule[1.1pt]
\textbf{Judger} & \textbf{QuixBugs (\%)} & \textbf{DebugBench (\%)}& \textbf{Tie (\%)}\\
\midrule[0.7pt]
GPT-4 & 42.6 & 57.1 & 0.3\\
Claude 3 & 46.9 & 53.1 & 0.0\\
\bottomrule[1.1pt]
\end{tabular}
}
\caption{Win Rate of comparison on bugs likely to occur in real-world scenarios as judged by GPT-4~\citep{openai2023gpt4} and Claude 3 Sonnet~\citep{anthropic2024claude} with instruction "Which bug is likely to occur in real-world scenarios?". }
\label{tab:bug_comparison}
\end{table}

Based on these observations, we assert that the bugs in DebugBench, despite being synthetically injected by large models, do not exhibit significant differences from those found in real-world scenarios. In this particular evaluation scenario, it even surpasses QuixBugs in terms of the realism of bug instances. Therefore, we believe that the aforementioned concerns are substantially mitigated.

\begin{table*}[!t]
\centering
\scalebox{0.75}{
\begin{tabular}{cc|cccc|cc|c}
\toprule[1.1pt]
\textbf{Major Category} & 
\textbf{Minor Type} & 
\codefont{CodeLlama}&
\codefont{Llama-3}&
\codefont{DeepSeek}&
\codefont{Mixtral}&
\codefont{gpt-3.5}&
\codefont{gpt-4}&
human\\ 
\midrule[0.7pt]
\multirow{7}{*}{Syntax}& 
misused ==/=& 
18.2&
58.4&
68.6&
12.4&
70.5&
87.9&
11/12\\
&missing colons& 
23.3&
44.2&
62.8&
25.6&
80.9&
93.6&
12/12\\
&unclosed parentheses& 
27.1&
51.9&
86.5&
14.3&
81.2&
89.6&
12/12\\
&illegal separation& 
7.4&
61.8&
77.9&
17.6&
78.1&
89.0&
12/12\\
&illegal indentation& 
4.4&
42.2&
77.8&
28.9&
79.6&
87.8&
12/12\\
&unclosed string& 
28.8&
48.0&
94.4&
9.6&
82.0&
91.4&
12/12\\
&illegal comment& 
31.5&
41.1&
45.2&
12.9&
67.4&
78.0&
11/12\\
\midrule[0.7pt]
\multirow{4}{*}{Reference}& 
faulty indexing& 
27.2&
53.4&
67.5&
11.7&
72.9&
77.1&
10/12\\
&undefined objects& 
21.9&
54.5&
68.4&
4.3&
70.6&
81.7&
12/12\\
&undefined methods& 
15.0&
46.7&
43.7&
6.6&
59.3&
78.5&
11/12\\
&illegal keywords& 
58.1&
13.5&
57.3&
18.5&
76.1&
83.6&
11/12\\
\midrule[0.7pt]
\multirow{4}{*}{Logic}& 
condition error& 
13.5&
46.5&
47.7&
22.3&
58.5&
73.1&
10/12\\
&operation error& 
8.3&
28.3&
27.8&
3.3&
49.5&
68.6&
10/12\\
&variable error& 
10.0&
29.0&
38.0&
10.0&
52.3&
63.1&
9/12\\
&other error& 
8.0&
40.0&
44.0&
2.0&
61.1&
72.2&
10/12\\
\midrule[0.7pt]
\multirow{3}{*}{Multiple}& 
double bugs& 
3.3&
43.2&
46.1&
8.4&
56.4&
70.7&
11/12\\
&triple bugs& 
6.7&
29.3&
54.5&
5.6&
45.5&
58.9&
9/12\\
&quadraple bugs& 
5.0&
31.2&
49.2&
4.5&
38.7&
55.9&
8/12\\

\bottomrule[1.1pt]
\end{tabular}} 
\caption{Debugging performance of various models against human proficiency measured by Pass Rate. Model names are abbreviated for clarity: \codefont{CodeLlama} represents \codefont{CodeLlama-7b-Instruct}; \codefont{Llama-3} is short for \codefont{Llama-3-8B-Instruct}; \codefont{DeepSeek} is short for \codefont{DeepSeek-Coder-33B-Instruct}; \codefont{Mixtral} represents \codefont{Mixtral-8x7B-Instruct}; \codefont{gpt-3.5} denotes \codefont{gpt-3.5-turbo-0613}; and \codefont{gpt-4} refers to \codefont{gpt-4-0613}. The experimental results reveal that closed-source models are less effective compared to human performance and open-source models attains even lower Pass Rate scores in debugging tasks.}
\label{tab:comparative debugging}
\end{table*}

\section{Experiments}

\paragraph{Evaluated Models.}
To obtain a comprehensive understanding of LLMs' debugging capabilities and identify the potential gap between open-source and closed-source models, we conduct experiments on two popular commercial models: \codefont{gpt-3.5-turbo-0613}~\citep{openai2022chatgpt} and \codefont{gpt-4-0613}~\citep{openai2023gpt4}. For open-source models, we select \codefont{CodeLlama-7b-Instruct}~\citep{rozière2023code}, \codefont{Llama-3-8B-Instruct}~\citep{llama3_website}, \codefont{DeepSeek-Coder-33B-Instrct}~\citep{deepseek-coder} and \codefont{Mixtral-8x7B-Instruct}~\citep{jiang2024mixtral} for assessment.

\paragraph{Metric.}

The metric for DebugBench is based on the test suites\footnote{Users of the benchmark will require a LeetCode account to access these test suites.} provided by \citet{leetcode2023}. These suites include a mix of 1-3 known test cases and 8-100 unknown test cases for each instance. 

Specifically, we use Pass Rate to quantify the debug ability of language models. For a buggy snippet $\theta_i$ and its fixed version $\theta_i^*$, we have a corresponding set of test cases $(x_i^{0}, y_i^{0}), (x_i^{1}, y_i^{1}), ..., (x_i^{m}, y_i^{m})$. Whether the bug instance is successfully repaired can be referred to as $\bigwedge_{j=0}^{m} [a_{\theta_i^*}(x_i^{j}) = y_i^{j}]
$, an aggregate result of all test cases. The Pass Rate, $PR$, that represents the test result on $n$ bug instances are defined as:
\begin{align}
PR = \sum^{n}_{i=0}\frac{\bigwedge_{j=0}^{m} [a_{\theta_i^*}(x_i^{j}) = y_i^{j}]
}{n}\times100\%\nonumber
\end{align}

\paragraph{Human Performance.}

The proficiency of human debuggers is assessed by three programmers, each with over four years of experience in programming. Before the formal experiment, they underwent a two-hour training session focused on understanding the purpose of human evaluation and the criteria for metrics. This was followed by a one-hour trial session. Each participant independently debugged 72 bugs, dedicating approximately 20 hours per person. During this process, access to Integrated Development Environments (IDEs) was provided to facilitate runtime analysis but any access to deep learning tools like \citet{github_copilot} was prohibited.

\subsection{Debugging Capabilities}
We evaluate the debugging capabilities of LLMs by assessing two closed-source and four open-source LLMs across 18 types of programming errors in three distinct scenarios. 


\subsubsection{Overall Results}

\paragraph{Close-Source Models}
As shown in Figure \ref{fig:radar} and Table \ref{tab:comparative debugging}, we examined the performance of closed-source models, \codefont{gpt-4-0613}~\citep{openai2023gpt4} and \codefont{gpt-3.5-0613}~\citep{openai2022chatgpt}. They respectively pass 75.0\% and 62.1\% of the bug instances, achieving a level of debugging performance below human. The superiority of human debuggers can be attributed to robust test cases and interaction with the program through breakpoints and developmental environments. Despite LLMs' limited effectiveness, they exhibit significant time efficiency. The models complete inference processes for one bug in less than 10 seconds, a task that averagely takes humans around 20 minutes. This indicates that commercial models are now capable of partially achieving the objectives of Automated Debugging, bringing benefits in time efficiency, cost reduction, and minimizing human labor. The zero-shot prompts utilized in model evaluation can be found in Appendix \ref{sec:prompts}. 

\paragraph{Open-Source Models}
As illustrated in Table \ref{tab:comparative debugging}, most open-source models attain relatively lower Pass Rate score than commercial models. This underscores a notable shortfall in the zero-shot debugging abilities of open versus closed-source LLMs. The ineffectiveness is likely due to a limited presence of debugging data in their pre-training datasets. These findings highlight the need for an open-source model capable of supporting debugging for research utility and practical applications. However, it is also worth noting that some open-source models like \codefont{DeepSeek-Coder-33B-Instrct} do achieve a level that is close to commercial models. It is also worth noticing that some models like \codefont{Mixtral-8x7B-Instruct} fail certain cases because they cannot comply with the instructions regarding the format.

\begin{figure}[!t]
  \centering
  \includegraphics[width=0.5\textwidth]{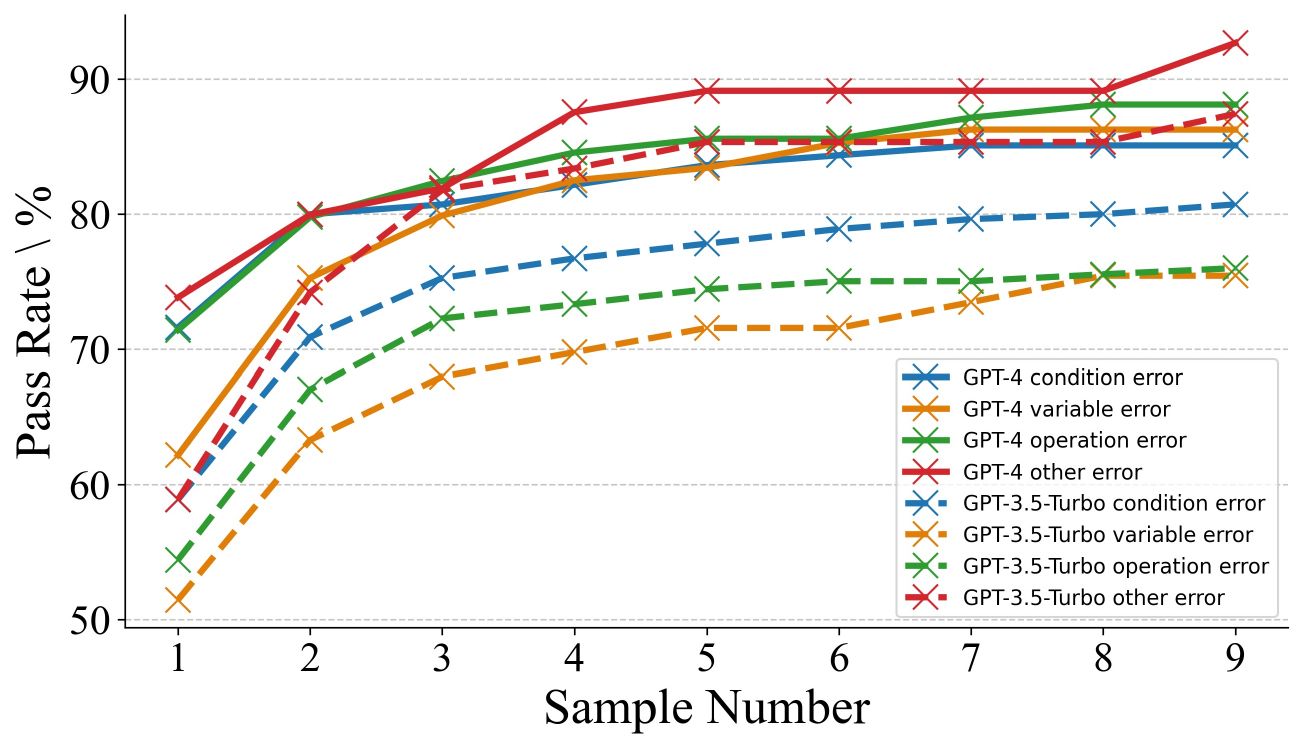}
  \caption{Pass Rate of GPT-4 \citep{openai2023gpt4} and GPT-3.5-turbo \citep{openai2022chatgpt} with more samples containing logical errors, particularly noting a significant improvement from 1 to 4 samples.}
  \label{fig:multi_sample}
\end{figure}

\subsubsection{Effect of Bug Types}
As illustrated in Figure \ref{fig:radar} and Table \ref{tab:comparative debugging}, the challenge of debugging varies markedly with the bug type for both humans and models. Syntax and reference errors are comparatively simpler to spot and rectify. In contrast, logic bugs pose a greater challenge, requiring an understanding of the code's underlying mechanisms. Additionally, the complexity of debugging escalates with an increase in the number of bugs within a code snippet. Therefore, in training or improving models for debugging, special emphasis should be placed on enhancing their ability to handle logic errors and scenarios with multiple concurrent errors.

\subsubsection{In-depth Analysis}
In this section, we examine two additional scenarios for deeper analyses.

\begin{figure}[!t]
  \centering
  \includegraphics[width=0.5\textwidth]{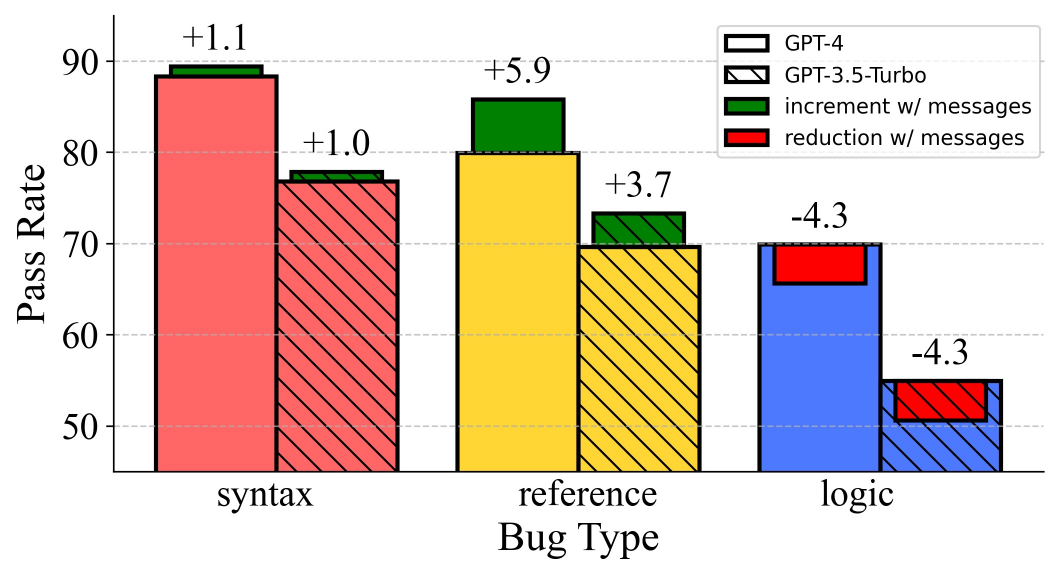}
  \caption{Effect of runtime feedback on \codefont{gpt-4-0613} \citep{openai2023gpt4} and \codefont{gpt-3.5-turbo-0613}'s \citep{openai2022chatgpt} debugging performance. It improves syntax and reference error handling but impairs logic error resolution.}
  \label{fig:runtime}
\end{figure}

\paragraph{Effect of Multiple Sampling.}
In this scenario, a language model is permitted to generate multiple responses to a single debugging query. An instance is marked as `pass' if at least one response successfully meets all test case criteria. Due to budget constraints, we limit our sampling to a maximum of nine answers for each instance with logical errors. As illustrated in Figure \ref{fig:multi_sample}, increased sampling attempts enhance debugging performance, indicating an effective trade-off: better debugging at the cost of using more inference tokens.

\paragraph{Effect of Runtime Feedback.}

Recent studies~\citep{chen2023teaching,jiang2023selfevolve} find out that providing runtime information like program output and traceback messages enhances the coding capabilities of LLMs. In this section, we investigate the influence of runtime messages on the debugging process. We leveraged the built-in runtime environment of the LeetCode test suites to obtain feedback information. As illustrated in Figure \ref{fig:runtime}, the runtime feedback has a clear impact on the debugging performance of LLMs. For syntax and reference errors, traceback information effectively identifies the locations of bugs, thereby facilitating the debugging process. However, for logic bugs, the details provided in traceback messages are often too low-level to facilitate effective debugging and may even cause disruptions. This indicates that the information provided by Runtime Feedback is not always useful for debugging LLMs.
Positive and negative examples of runtime messages are accessible in Appendix \ref{sec:Runtime Messages}.

\subsection{Interplay between Debugging and Coding}
As an extension of the evaluation of debugging capabilities, we compare the difficulty and correlation of these tasks to deepen our understanding of LLMs' proficiency in both code generation and debugging.



\begin{figure}[!t]
  \centering
  \includegraphics[width=0.5\textwidth]{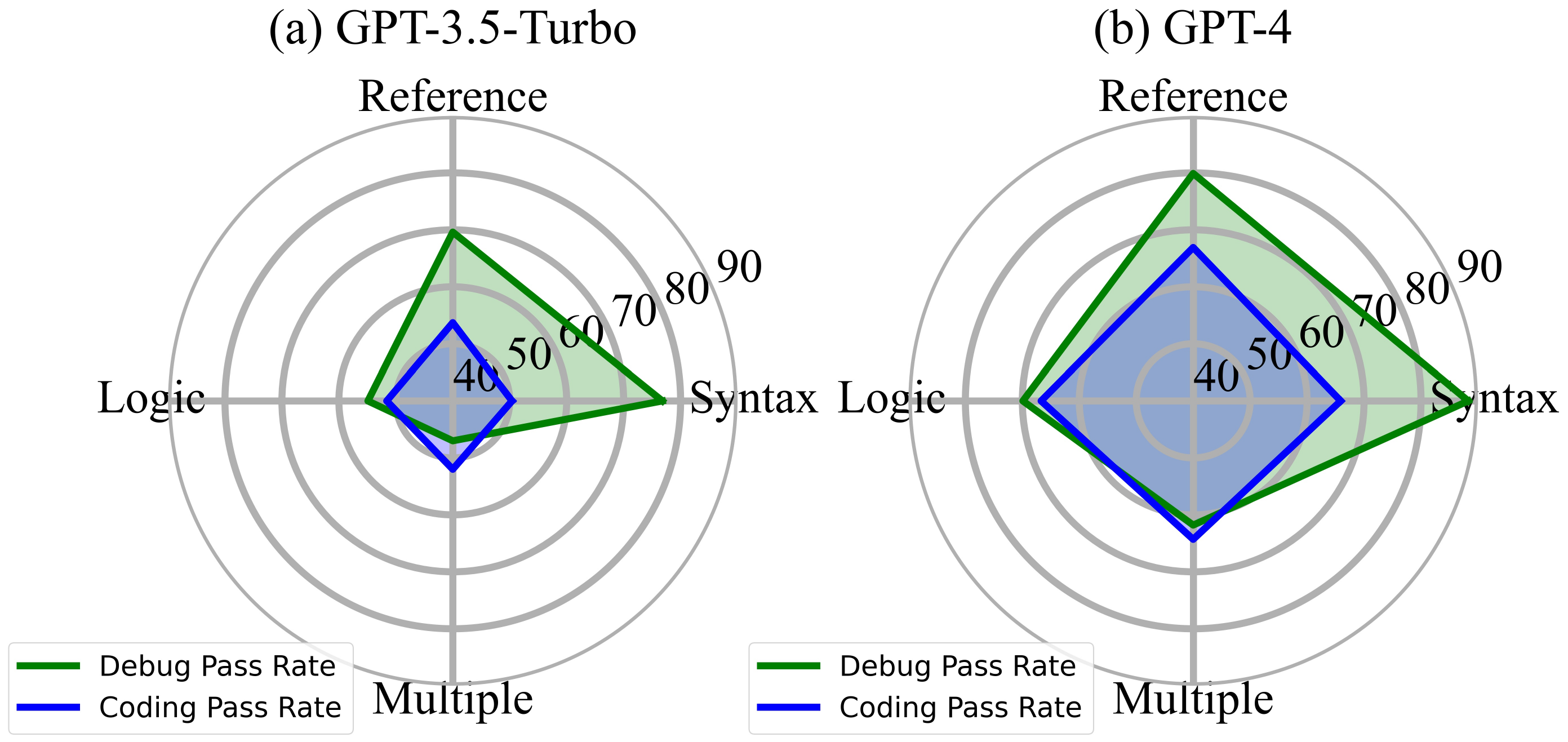}
  \caption{Pass Rate of coding and debugging tasks with same programming problems. }
  \label{fig:difficult}
\end{figure}

\paragraph{Comparison of Difficulty.}
We analyze the debugging and code generation performance of \codefont{gpt-4-0613}\citep{openai2022chatgpt} and \codefont{gpt-3.5-turbo-0613}\citep{openai2023gpt4} on identical instances. As illustrated in Figure \ref{fig:difficult}, we find that correcting syntax and reference errors typically presents less difficulty than generating full code for a specific query, while addressing logical errors or multiple issues can be as challenging as code generation itself. This pattern implies that the task of debugging is relatively easier than code generation.

\paragraph{Correlation between Debugging and Coding}
To explore the correlation between debugging and coding, i.e., whether a programming question is more likely to be easy-to-debug if it is easy-to-code and vice versa, we compute the Phi-Coefficient for closed-source LLMs and find that all categories of bugs have a positive Phi-Coefficient score with code generation ranging from 0.1 to 0.3 as shown in Table \ref{tab:phi_coefficient}. This suggests that the capabilities of LLMs to approach these two tasks are explicitly correlated. 



\begin{table}[htbp]
\centering
\small
\begin{tabular}{lcc}
\toprule[1.1pt]
\textbf{Model} & \textbf{Bug Type} & \textbf{Phi-Coefficient}\\
\midrule[0.9pt]
\multirow{4}{*}{GPT-4} & syntax & 0.221\\
& reference & 0.115 \\
& logic & 0.353 \\
& multiple & 0.273 \\
\midrule[0.9pt]
\multirow{4}{*}{GPT-3.5-Turbo} & syntax & 0.148\\
& reference & 0.196 \\
& logic & 0.174 \\
& multiple & 0.298 \\
\bottomrule[1.1pt]
\end{tabular}
\caption{Phi-Coefficient of LLMs' coding and debugging performance.}
\label{tab:phi_coefficient}
\end{table}

\section{Related Work}

\subsection{LLM-based Coding}
\label{sec:LLM in Code Generation}
The field of LLM code generation has been extensively studied. Researchers have collected code corpora to train large language models that specialize in code generation~\citep{chen2021evaluating,nijkamp2022codegen,li2023starcoder}. General-purpose LLMs also demonstrated impressive coding abilities as a result of extensive pre-training on datasets rich in code-related content~\citep{touvron2023llama, workshop2022bloom, openai2022chatgpt, openai2023gpt4}. Parallel to the development of these foundational models, innovative methods such as verbal reinforcement learning with feedback from runtime messages \citep{shinn2023reflexion}, and multi-agent collaboration \citep{qian2023communicative}, have been implemented to further refine the coding abilities of LLMs.

As another key component of programming proficiency, LLMs' debugging capabilities have not garnered so much attention. This can be partly attributed to the absence of evaluation benchmarks. To overcome this deficiency, we introduce DebugBench, the new LLM debugging benchmark discussed in this work.

\subsection{Automated Program Repair}
\label{sec:Automated Program Repair}
Automated Program Repair (APR) refers to the process of automatically fixing program bugs or errors without human intervention. This topic has gained significant attention due to its potential to reduce the time and cost in software development \citep{goues2019automated}. While template-based \citep{liu2019tbar}, search-based \citep{ke2015repairing} and generic \citep{le2011genprog} methods have been proposed to solve the task, program repair based on Large Language Models exhibit significant potential \citep{prenner2022can}.

\citet{prenner2022can} evaluated OpenAI's CodeX \citep{chen2021evaluating} on QuixBugs \citep{lin2017quixbugs} and found LLM debugging promising. \citet{sobania2023analysis} utilized ChatGPT \citep{openai2022chatgpt} to address bugs in QuixBugs, outperforming the previous state-of-the-art. \citet{xia2023conversational} tested LLM debugging with a conversational strategy to refine the debugging patches based on the feedback from each turn on QuixBugs and achieved higher performance. However, \citet{zhang2023critical} pointed out that evaluations on traditional APR datasets like QuixBugs \citep{lin2017quixbugs} and Defects4J \citep{just2014defects4j} face a severe risk of data leakage and evaluate ChatGPT~\citep{chatgpt2023} on a new benchmark of 151 bugs from competitive programming problems.

These works are fundamental in verifying the feasibility of LLMs for debugging, but they face challenges that require further investigation. Apart from data leakage, current evaluations of LLM debugging face significant constraints: limited bug diversity and constrained test scale as illustrated in Table \ref{tab:evaluation_comparison}. In order to gain a deeper understanding of the potential for LLM debugging, we conducted a systematic evaluation with \datasetname.

\section{Conclusion}
In this work, we presented \datasetname, a benchmark specifically designed to evaluate the debugging capabilities of large language models. \datasetname was developed utilizing source data from LeetCode \citep{leetcode2023} and bug implantation with prompted GPT-4, underpinned by a stringent quality control process.

Our experiments with \datasetname revealed several key findings: (1) In a zero-shot scenario, closed-source models exhibited relatively lower debugging performance compared to humans, while open-source models attained a even lower pass rate score in zero-shot scenarios; (2) Multiple and logical errors posed a greater challenge for the repair compared to syntax and reference errors; (3) Runtime feedback enhanced the debugging performance for syntax and reference errors but is unhelpful for logical errors; (4) For closed-source LLMs, debugging syntax or reference errors is easier than code generation while logic or multiple errors can equally hard or even harder. And their capabilities in coding and debugging are correlated. Hopefully, these findings will contribute to the advancement of large language models in the field of automatic debugging. The data is open-sourced via Apache-2.0 license.

\paragraph{Future Work}
 The scope of debugging scenarios can be expanded to more practical and complex situations like repository-level debugging \citep{jimenez2023swe, bairi2023codeplan} and scenarios involving human interaction. Additionally, based on the results of human evaluation, the ability to write reliable test cases and interact with Integrated Development Environments (IDEs) significantly boosts manual debugging performance. It can be meaningful to evaluate how well LLMs write test cases and interact with IDEs for debugging.

\section*{Limitations}
This study has certain limitations that must be acknowledged. The bugs in our experiments were synthetically created and might not entirely reflect the intricacies of real-world coding scenarios. The scope of our study was confined to four open-source and two closed-source models, which do not represent the full spectrum of existing LLMs. 

Another limitation of this benchmark is the evaluation metric, which is based on test cases from the LeetCode platform. While the test suites used in the benchmark are of high quality, they may not be as accessible as those local test cases. In particular, the benchmark requires testers to have one or more LeetCode accounts in order to run the online judge system, and testers may be impacted by rate limits and query limits imposed by the LeetCode website. 

\section*{Acknowledgements}
This work is supported by the National Key R\&D Program of China (No.2022ZD0116312),Quan Cheng Laboratory (Grant No. QCLZD202301), the Postdoctoral Fellowship Program of CPSE (Grant No. GZB20230343), China Postdoctoral Science Foundation (Grant No. 2023M741945), and Institute Guo Qiang at Tsinghua University.

\bibliography{custom}

\clearpage
\appendix

\section*{Appendix}
\section{Prompts}
\label{sec:prompts}
In this section, we detail the prompt utilized in the process of bug implantation and model evaluation.

The following is the prompt we use for GPT-4 to implant different types of bugs into code snippets.
\begin{prompt}
\textbf{Prompt: }
Observe the following code implementation. Your task is to first add a(n) \{\codefont{BUG\_TYPE}\} bug to the code and then to explain the bug you add in 15 words. You have to write the implementation again. You should put <bug> </bug>, <exp> </exp> in the beginning and end of the code and explanation. Make sure the bug incurs unexpected behaviors. Do not write anything else in your response. You must not add any comment about the bug in code.
\{\codefont{CODE IMPLEMENTATION}\}
\end{prompt}

\vspace{8pt}
The following is the prompt for zero-shot evaluation.
\begin{prompt}
\textbf{Prompt: }Observe the following faulty code. Your task is to fix up the code and explain on the modification in less than 20 words. You have to write the fixed code again. You should put <code></code> and <exp></exp> on the boundary of the code and the explanation. Do not write anything else in your response. 
\{\codefont{BUG INSTANCE}\}
\end{prompt}

\vspace{8pt}
This is the simplified version of zero-shot prompt.
\begin{prompt}
\textbf{Simplified Prompt: }Observe the following faulty code implementation. Your task is to fix up the code.
\{\codefont{BUG INSTANCE}\}
\end{prompt}

\vspace{8pt}
Below is the three shot prompt for model evaluation.
\begin{prompt}
\textbf{Three-Shot Prompt: }Observe the following faulty code implementation. Your task is to fix up the code.

Example 1 \{\codefont{BUG INSTANCE1}\} 

\{\codefont{FIXED INSTANCE1}\}

Example 2 ...

Example 4 \{\codefont{BUG INSTANCE4}\}

\{\codefont{FIXED INSTANCE4}\}

Test Example \{\codefont{BUG INSTANCE}\}
\end{prompt}

\section{Bug Type Definition}
\label{sec:Definition and Distribution of Each Type of Bugs}
The bug instances in DebugBench are categorized into four major categories and 18 minor types. The definitions and distributions of these categories are detailed in Table \ref{tab:sub_type_numbers} and Table \ref{tab:subtype_numbers}. It's important to note that instances of multiple bugs constitute a significant portion of the total count. This is because they can be further classified according to specific combinations of bug types, rather than simply by the number of bugs. To ensure a sufficient number of instances for each unique combination of bugs, a substantial volume of instances is required.

\begin{table}[htbp]
\centering
\begin{tabular}{lc}
\toprule
\textbf{Type} & \textbf{Type Number} \\
\midrule
Syntax Error& 761 \\
Reference Error& 684 \\
Logic Error& 590 \\
Multiple Error& 2218 \\
\bottomrule
\end{tabular}
\caption{Distribution of major bug types.}
\label{tab:sub_type_numbers}
\end{table}

\begin{table*}[htbp]
\centering
\begin{tabular}{cp{8cm}c}
\toprule
\textbf{Minor Types} & \textbf{Definition} & \textbf{Instance Number} \\
\midrule
misused ==/= &  Misuse between equality (==) or assignment (=) operators. & 137 \\
missing colons & Omission of colons where required, such as in control structures (if, for, while) or function definitions in certain programming languages. & 129 \\
unclosed parentheses & Failure to close a set of parentheses, leading to syntax errors. & 133 \\
illegal separation & Improper use of separators like commas or semicolons, causing syntax errors. & 68 \\
illegal indentation & Incorrect indentation that violates the syntax rules of indentation-sensitive languages like Python. & 45 \\
unclosed string & A string literal that is not properly closed with matching quotation marks. & 125 \\
illegal comment & Use of incorrect syntax for comments, or placing comments where they are not allowed. & 124 \\
\midrule
faulty indexing & Accessing elements of a collection (like arrays or lists) with an incorrect index, often leading to runtime errors. & 206 \\
undefined objects & Reference to an object that has not been defined or imported. & 187 \\
undefined methods & Calling a method that doesn't exist for a given object or class. & 167 \\
illegal keywords & Misuse of reserved words in a programming language. & 124 \\
\midrule
condition error & Logical errors in conditions used in control structures. & 260 \\
operation error & Errors in arithmetic or other operations, such as division by zero. & 180 \\
variable error & Errors related to variable misuse, like using an uninitialized variable. & 100 \\
other error & Any programming error that does not fit into the above categories. & 50 \\
\midrule
double bug & Two distinct bugs present in a single snippet or section of code. & 750 \\
triple bug & Three distinct bugs present in a single snippet or section of code. & 750 \\
quadruple bug & Four distinct bugs present in a single snippet or section of code. & 718 \\
\bottomrule
\end{tabular}

\caption{Definition and distribution of each minor type of bugs from categories of syntax error, reference error, logic error and multiple error.}
\label{tab:subtype_numbers}
\end{table*}

\section{Instance Feature Schema}
\label{sec:feature schema}
As exhibited in Figure \ref{fig:construction}, the features of an instance in \datasetname includes `Question', `Example', `Oracle Code Solution', `Buggy code', and `Bug Explanation'. 
`Question' refers to the original programming query, which decides the requirement for the program. `Example' encompasses one to three input-output pairs, which serve either as supplementary explanations of the program requirements or as test cases for debugging. `Oracle Code Solution' denotes the accurate code implementation that successfully passes all test cases, acting as the annotation for correct solutions. `Buggy Code' consists of code snippets embedded with one or more errors, forming the primary subject for debugging processes. Lastly, `Bug Explanation' provides a brief overview by GPT-4 \citep{openai2023gpt4} regarding the nature of the bug, which aids in understanding the bug and and quality control. Qualitative examples are accessible in Appendix \ref{sec:Examples of Synthesized Bugs}.

\section{Expenditure Estimation}
\label{sec:Expenditure Estimation}

We employed commercial models from OpenAI for constructing and evaluating our dataset which cost around \$330. The estimated expenditures for this process are detailed in Table \ref{tab:expenditure approximation}.

\begin{table*}[htbp]
\centering
\scalebox{0.85}{
\begin{tabular}{lccccc}
\toprule[1.1pt]
\textbf{Process} & \textbf{Query Number} & \textbf{Token Num} & \textbf{Price} & \textbf{Model Price} & \textbf{Expenditure} \\
\midrule[0.7pt]
Bug Implantation & 2760 & 500 / 500 & \$0.03 / \$0.06 (1k tokens) & gpt-4 & \$124.2 \\
Model Evaluation & 4405 & 500 / 500 & \$0.03 / \$0.06 (1k tokens) & gpt-4 & \$198.2 \\
Model Evaluation & 4405 & 500 / 500 & \$0.0010 / \$0.0020 (1k tokens) & gpt-3.5-turbo & \$6.6 \\
\bottomrule[1.1pt]
\end{tabular}}
\caption{Expenditure Approximation of Commercial Models. Token number and model price are written in a format of input token / output token.}
\label{tab:expenditure approximation}
\end{table*}

\section{Cases Failing Manual Inspection}
\label{sec:Cases Failing Manual Inspection.}

During manual inspection, we propose three evaluating criteria (1) The bugs must cause the intended malfunction, fail specific test cases, and align with the assigned bug type and description. (2) Sensitive Information Security: The instances must be devoid of sensitive data, such as personal information. (3) Scenario Alignment: The bugs should resemble those found in actual code debugging scenarios and should not include obvious clues, like comments indicating the bug’s location.

We detect some instances that violate code (1) and code (3) and showcase them as follows.\\

\paragraph{Wrong Categories.} GPT-4~\citep{openai2023gpt4} occasionally introduces bugs that differ from their assigned category, thereby violating code principle (1). An example of this is a bug generated in response to a prompt specifying 'undefined methods.' Rather than invoking undefined functions as expected, the code triggers an infinite loop.

\begin{prompt_red}
\textbf{Buggy Code: }
 \begin{lstlisting}
class TreeAncestor {
public:
    vector<vector<int>>v;
    TreeAncestor(int n, vector<int>& parent) {
        vector<vector<int>> par(n, vector<int>(20));
        for (int i = 0; i < n; i++) par[i][0] = parent[i];
        for (int j = 1; j < 20; j++) {
            for (int i = 0; i < n; i++) {
                if (par[i][j - 1] == -1) par[i][j] = -1;
                else par[i][j] = par[par[i][j - 1]][j - 1];
            }
        }
        swap(v, par);
    }
    int getKthAncestor(int node, int k) {
        for (int i = 0; i < 20; i++) {
            if ((k >> i) & 1) {
                node = v[node][i];
                if (node == -1) return -1;
                getKthAncestor(node, k);
            }
        }
        return node;
    }
};
\end{lstlisting}
\end{prompt_red}

\paragraph{Unlikely Bugs.}  Some instances of bugs are accurately categorized by their error type, yet they are unlikely to appear in practical scenarios. For instance, it is atypical for 'undefined objects' in normal circumstances to be named 'undefined', which violates code (3).

\begin{prompt_red}
\textbf{Buggy Code: }
 \begin{lstlisting}
class Solution {
public:
    bool doesValidArrayExist(vector<int>& derived) {
        int xoor = 0;
        vector<int>& undefined;
        for (auto i: derived) xoor ^= i;
        xoor ^= undefined[0];
        return !xoor;
    }
};
\end{lstlisting}
\end{prompt_red}

\section{Runtime Messages}
\label{sec:Runtime Messages}

In this section, we present several examples of runtime messages that may or may not provide clues for the debugging process.

When dealing with bugs that raise runtime errors, these information are specifically helpful. For example, the following runtime messages directly points out the bug.
\begin{prompt_red}
\textbf{Buggy Code:}
 \begin{lstlisting}
 class Solution:
    def dp(self,i,s,prev,k,ct,n,dct:
        if k<0:
            return float("infinity")
        if i>=n:
            x=0
            if ct>1:
                x=len(str(ct))+1
            elif ct==1:
                x=1
            return x
        if (i,prev,ct,k) in dct:
            return dct[(i,prev,ct,k)]
        if s[i]==prev:
            inc=self.dp(i+1,s,prev,k,ct+1,n,dct)
        else:
            x=0
            if ct>1:
                x=len(str(ct))+1
            elif ct==1:
                x=1
            inc=x+self.dp(i+1,s,s[i],k,1,n,dct)
        exc=self.dp(i+1,s,prev,k-1,ct,n,dct)
        dct[(i,prev,ct,k)]=min(inc,exc)
        return min(inc,exc)


    def getLengthOfOptimalCompression(self, s: str, k: int) -> int:
        n=len(s)
        return self.dp(0,s,"",k,0,n,{})
 \end{lstlisting}
\end{prompt_red}

\begin{prompt_red}
\textbf{Runtime Messages: }

\codefont{Line 3: SyntaxError: '(' was never closed}
\end{prompt_red}

\vspace{8pt}
Traceback messages may sometimes be too low-level to offer effective information. The following bug change one detail of operation about the prime arrangements from '+n' to '-n', for which the 'stackoverflow' messages does not provide any help.
\begin{prompt_red}
\textbf{Buggy Code:}
 \begin{lstlisting}
class Solution {
public:
    long long fact(int n)
    {
        if(n<=1)return 1;
        return (n*fact(n+1)%1000000007)%1000000007; 
    }
    int numPrimeArrangements(int n) {
        if(n==1)return 1;
        if(n<=3)return n-1;
        int t=0,flag;
        for(int i=2;i<=n;i++)
        {
            flag=0;
            for(int j=2;j<=sqrt(i);j++)
            {
                if(i%j==0)
                {
                    flag=1;
                    break;
                }
            }
            if(flag==0)
            {
                t++;
            }
        }
        return (fact(t)*fact(n-t))%1000000007;

    }
};
 \end{lstlisting}
\end{prompt_red}

\begin{prompt_red}
\textbf{Runtime Messages: }

\codefont{AddressSanitizer: stack-overflow on the address 0x7ffcdc8e9ff8 (pc 0x000000366e14 bp 0x7ffcdc8ea000 sp 0x7ffcdc8ea000 T0)}
\end{prompt_red}

\section{Comparison of Bug Realism}
\label{sec: Comparison of Bug Realism}
To address concerns regarding the realism of synthetically generated bugs, we conducted a comparative study between instances in DebugBench and those from real-world datasets. Specifically, we examined QuixBugs~\citep{lin2017quixbugs}, which encompasses bugs derived from the Quixey Challenge and based on real-world scenarios. In this study, we compared the realism of all 40 Python and Java bugs from QuixBugs with corresponding instances in DebugBench of the same bug type and programming language. Each case was compared against five random cases from DebugBench, with the order switched to mitigate bias.

As judges, we employed GPT-4~\citep{openai2023gpt4} and Claude 3 Sonnet~\citep{anthropic2024claude} to perform the evaluations. We set the temperature parameter to $0.01$ and top\_p to $0.7$. The prompt used for this evaluation is detailed below.

\begin{prompt_blue}
\textbf{System Prompt: }

\codefont{You are a helpful assistant with profound knowledge in coding. Here are two pieces of buggy code and explanations about their bugs. Your job is to judge which one is more likely to occur in realistic scenarios. You have three output options: <Code Piece A>, <Code Piece B>, and <About the same>. You should only output an option and you should not output anything else.}
\end{prompt_blue}

\begin{prompt_blue}
\textbf{User Message: }

\codefont{\# Code Piece A

...

\# Explanation about Code Piece A

...

\# Code Piece B

...

\# Explanation about Code Piece B

...}
\end{prompt_blue}

\quad

The results of this study are presented in Table~\ref{tab:bug_comparison}. As illustrated, the bug instances in DebugBench are considered to have similar realism to those in QuixBugs. Consequently, we believe that concerns regarding the realism of the bugs in DebugBench can be effectively mitigated.

\section{Examples in BuGalley}
\label{sec:Examples of Synthesized Bugs}
In this section we demonstrated 4 positive and 4 negative cases of source data (blue), buggy code (red), LLM generation and evaluation results (green) of different language, different models and different types of bugs. \\
\vspace{10pt}

\noindent\textbf{1. GPT-4, Python3, colon missing}

\begin{prompt_blue}
\textbf{Question: } Given a sentence text (A sentence is a string of space-separated words) in the following format:

First letter is in upper case.
Each word in text is separated by a single space.

Your task is to rearrange the words in text such that all words are rearranged in an increasing order of their lengths. If two words have the same length, arrange them in their original order.
Return the new text following the format shown above.
\end{prompt_blue}

\begin{prompt_blue}
\textbf{Examples: } Input: text = "Leetcode is cool"\\Output: "Is cool leetcode"\\Explanation: There are 3 words, "Leetcode" of length 8, "is" of length 2 and "cool" of length 4.\\Output is ordered by length and the new first word starts with capital letter.
\end{prompt_blue}

\begin{prompt_blue}
\textbf{Oracle Solution Code: }
\begin{lstlisting}
class Solution:
    def arrangeWords(self, text: str) -> str:
        return " ".join(sorted(text.split(), key=len)).capitalize()
\end{lstlisting}
\end{prompt_blue}

\begin{prompt_red}
\textbf{Buggy Code: }
\begin{lstlisting}
class Solution:
    def arrangeWords(self, text: str) -> str
        return " ".join(sorted(text.split(), key=len)).capitalize()
\end{lstlisting}
\end{prompt_red}

\begin{prompt_red}
\textbf{Bug Explanation: } A colon is missing at the end of the function definition line.
\end{prompt_red}

\begin{prompt_green}
\textbf{GPT-4 Debugging: }
\begin{lstlisting}
class Solution:
    def arrangeWords(self, text: str) -> str:
        return " ".join(sorted(text.split(), key=len)).capitalize()
\end{lstlisting}
\end{prompt_green}

\begin{prompt_green}
\textbf{Debugging Explanation: } The code was missing a colon at the end of the method definition.
\end{prompt_green}

\begin{prompt_green}
\textbf{Test Results: } 
Run time: 43 ms\\
Run Memory: 18624000\\
Test Cases Results: \seqsplit{11111111111111111111111111111111111111111111111111111111111111111111111111}
\end{prompt_green}

\begin{prompt_green}
\textbf{Test Decision: } 
True
\end{prompt_green}

\vspace{10pt}
\noindent\textbf{2. GPT-4, C++, undefined objects}

\begin{prompt_blue}
\textbf{Question: } A 0-indexed array derived with length n is derived by computing the bitwise XOR of adjacent values in a binary array original of length n.
Specifically, for each index i in the range [0, n - 1]:

If i = n - 1, then derived[i] = original[i] XOR original[0].
Otherwise, derived[i] = original[i] XOR original[i + 1].

Given an array derived, your task is to determine whether there exists a valid binary array original that could have formed derived.
Return true if such an array exists or false otherwise.

A binary array is an array containing only 0's and 1's
\end{prompt_blue}

\begin{prompt_blue}
\textbf{Examples: } Input: derived = [1,1,0]\\Output: true\\Explanation: A valid original array that gives derived is [0,1,0].\\derived[0] = original[0] XOR original[1] = 0 XOR 1 = 1 \\derived[1] = original[1] XOR original[2] = 1 XOR 0 = 1\\derived[2] = original[2] XOR original[0] = 0 XOR 0 = 0
\end{prompt_blue}

\begin{prompt_blue}
\textbf{Oracle Solution Code: }
\begin{lstlisting}
class Solution {
public:
    bool doesValidArrayExist(vector<int>& derived) {
        int xoor = 0;
        for (auto i: derived) xoor ^= i;
        return !xoor;
    }
};
\end{lstlisting}
\end{prompt_blue}

\begin{prompt_red}
\textbf{Buggy Code: }
\begin{lstlisting}
class Solution {
public:
    bool doesValidArrayExist(vector<int>& derived) {
        int xoor = 0;
        vector<int>& undefined;
        for (auto i: derived) xoor ^= i;
        xoor ^= undefined[0];
        return !xoor;
    }
};
\end{lstlisting}
\end{prompt_red}

\begin{prompt_red}
\textbf{Bug Explanation: } The bug is in `vector<int>\& undefined;` the vector undefined is declared but never defined.
\end{prompt_red}

\begin{prompt_green}
\textbf{GPT-4 Debugging: }
\begin{lstlisting}
class Solution {
public:
    bool doesValidArrayExist(vector<int>& derived, vector<int>& undefined) {
        int xoor = 0;
        for (auto i: derived) xoor ^= i;
        xoor ^= undefined[0];
        return !xoor;
    }
};
\end{lstlisting}
\end{prompt_green}

\begin{prompt_green}
\textbf{Debugging Explanation: } The undefined vector was declared but not defined and also was not part of function parameter.
\end{prompt_green}

\begin{prompt_green}
\textbf{Test Results: } 
Run time: N/A\\
Run Memory: 0\\
Test Cases Results: \seqsplit{compling error}
\end{prompt_green}

\begin{prompt_green}
\textbf{Test Decision: } 
False
\end{prompt_green}

\vspace{10pt}
\noindent\textbf{3. GPT-4, Java, operation error}

\begin{prompt_blue}
\textbf{Question: } Given two arrays arr1 and arr2, the elements of arr2 are distinct, and all elements in arr2 are also in arr1.
Sort the elements of arr1 such that the relative ordering of items in arr1 are the same as in arr2. Elements that do not appear in arr2 should be placed at the end of arr1 in ascending order.
\end{prompt_blue}

\begin{prompt_blue}
\textbf{Examples: } Input: arr1 = [2,3,1,3,2,4,6,7,9,2,19], arr2 = [2,1,4,3,9,6]\\Output: [2,2,2,1,4,3,3,9,6,7,19]
\end{prompt_blue}

\begin{prompt_blue}
\textbf{Oracle Solution Code: }
\begin{lstlisting}
import java.util.Arrays;

class Solution {
    public int[] relativeSortArray(int[] arr1, int[] arr2) {
        Arrays.sort(arr1);
        int[] ans = new int[arr1.length];
        int start = 0;
        int end = arr1.length - 1;
        int index = 0;

        for (int i = 0; i < arr2.length; i++) {
            int target = arr2[i];
            start = 0;
            end = arr1.length - 1;

            while (end >= start) {
                int mid = start + (end - start) / 2;
                if (arr1[mid] == target) {
                    for (int j = start; j <= end; j++) {
                        if (arr1[j] == target) {
                            ans[index++] = arr1[j];
                        }
                    }
                    break;
                }
                if (arr1[mid] < target) {
                    start = mid + 1;
                } else {
                    end = mid - 1;
                }
            }
        }

        for (int i = 0; i < arr1.length; i++) {
            if (index == arr1.length) {
                break;
            }
            boolean found = false;
            for (int num : arr2) {
                if (arr1[i] == num) {
                    found = true;
                    break;
                }
            }
            if (!found) {
                ans[index++] = arr1[i];
            }
        }

        return ans;
    }
}
\end{lstlisting}
\end{prompt_blue}

\begin{prompt_red}
\textbf{Buggy Code: }
\begin{lstlisting}
import java.util.Arrays;

class Solution {
    public int[] relativeSortArray(int[] arr1, int[] arr2) {
        Arrays.sort(arr1);
        int[] ans = new int[arr1.length];
        int start = 0;
        int end = arr1.length - 1;
        int index = 0;

        for (int i = 0; i < arr2.length; i++) {
            int target = arr2[i];
            start = 0;
            end = arr1.length - 1;

            while (end >= start) {
                int mid = start + (end - start) / 2;
                if (arr1[mid] == target) {
                    for (int j = start; j <= end; j++) {
                        if (arr1[j] == target) {
                            ans[index++] = arr1[j];
                            arr1[j] = -1;
                        }
                    }
                    break;
                }
                if (arr1[mid] < target) {
                    start = mid + 1;
                } else {
                    end = mid - 1;
                }
            }
        }

        Arrays.sort(arr1);

        for (int i = 0; i < arr1.length; i++) {
            if (index == arr1.length) {
                break;
            }
            boolean found = false;
            for (int num : arr2) {
                if (arr1[i] == num) {
                    found = true;
                    break;
                }
            }
            if (!found && arr1[i] != -1) {
                ans[index++] = arr1[i];
            }
        }

        return ans;
    }
}
\end{lstlisting}
\end{prompt_red}

\begin{prompt_red}
\textbf{Bug Explanation: } The bug I added was modifying the array in the sorting step without considering its impact on forthcoming operations, causing incorrect output.
\end{prompt_red}

\begin{prompt_green}
\textbf{GPT-4 Debugging: }
\begin{lstlisting}
import java.util.*;

class Solution {
    public int[] relativeSortArray(int[] arr1, int[] arr2) {
        int[] ans = new int[arr1.length];
        int index = 0;

        for (int i = 0; i < arr2.length; i++) {
            for (int j = 0; j < arr1.length; j++) {
                if (arr1[j] == arr2[i]) {
                    ans[index++] = arr1[j];
                }
            }
        }

        List < Integer > list = new ArrayList < > ();
        for (int i = 0; i < arr1.length; i++) {
            boolean found = false;
            for (int num : arr2) {
                if (arr1[i] == num) {
                    found = true;
                    break;
                }
            }
            if (!found) {
                list.add(arr1[i]);
            }
        }

        Collections.sort(list);

        for (Integer i: list) {
            ans[index++] = i;
        }

        return ans;
    }
}
\end{lstlisting}
\end{prompt_green}

\begin{prompt_green}
\textbf{Debugging Explanation: } Simplified logic to fill the resultant array by iterating through the second array and filling for each occurrence in the first array.
\end{prompt_green}

\begin{prompt_green}
\textbf{Test Results: } 
Run time: 4 ms\\
Run Memory: 40848000\\
Test Cases Results: \seqsplit{11111111111111111}
\end{prompt_green}

\begin{prompt_green}
\textbf{Test Decision: } 
True
\end{prompt_green}

\vspace{10pt}
\noindent\textbf{4. GPT-4, Python3, triple bugs}

\begin{prompt_blue}
\textbf{Question: } You are given the root of a binary tree and a positive integer k.
The level sum in the tree is the sum of the values of the nodes that are on the same level.
Return the kth largest level sum in the tree (not necessarily distinct). If there are fewer than k levels in the tree, return -1.
Note that two nodes are on the same level if they have the same distance from the root.
\end{prompt_blue}

\begin{prompt_blue}
\textbf{Examples: } Input: root = [5,8,9,2,1,3,7,4,6], k = 2\\Output: 13\\Explanation: The level sums are the following:\\- Level 1: 5.\\- Level 2: 8 + 9 = 17.\\- Level 3: 2 + 1 + 3 + 7 = 13.\\- Level 4: 4 + 6 = 10.\\The 2nd largest level sum is 13.
\end{prompt_blue}

\begin{prompt_blue}
\textbf{Oracle Solution Code: }
\begin{lstlisting}
# Definition for a binary tree node.
# class TreeNode:
#     def __init__(self, val=0, left=None, right=None):
#         self.val = val
#         self.left = left
#         self.right = right
class Solution:
    def kthLargestLevelSum(self, root: Optional[TreeNode], k: int) -> int:
        dq = collections.deque([root])
        a = []
        lvl = 1
        while dq:
            lvlsum = 0
            for i in range(len(dq)):
               n = dq.popleft()
               lvlsum += n.val
               if n.left: dq.append(n.left)
               if n.right: dq.append(n.right)
            a.append(lvlsum)
            lvl += 1
        a.sort(reverse=True)
        return a[k-1] if len(a) >= k else -1
\end{lstlisting}
\end{prompt_blue}

\begin{prompt_red}
\textbf{Buggy Code: }
\begin{lstlisting}
# Definition for a binary tree node.
# class TreeNode:
#     def __init__(self, val=0, left=None, right=None):
#         self.val = val
#         self.left = left
#         self.right = right
class Solution:
    def kthLargestLevelSum(self, root: Optional[TreeNode], k: int) -> int:
        dq = collections.deque([root])
        a = []
        lvl = 1
        while dq:
            lvlsum = 0
            for i in range(len(dq)):
               n = dq.popleft()
               lvlsum += n.val
               if n.left: dq.append(n.left)
               if n.right: dq.append(n.right)
                
            a.append(self.calculateLevelSum(lvlsum)) 
            lvl += 1
        a.sort(reverse=False)
        return a[k] if len(a) >= k else -1
\end{lstlisting}
\end{prompt_red}

\begin{prompt_red}
\textbf{Bug Explanation: } other error+ undefined methods+ indexing error
\end{prompt_red}

\begin{prompt_green}
\textbf{GPT-4 Debugging: }
\begin{lstlisting}
# Definition for a binary tree node.
# class TreeNode:
#     def __init__(self, val=0, left=None, right=None):
#         self.val = val
#         self.left = left
#         self.right = right
class Solution:
    def kthLargestLevelSum(self, root: TreeNode, k: int) -> int:
        dq = collections.deque([root])
        a = []
        while dq:
            lvlsum = 0
            for _ in range(len(dq)):
               n = dq.popleft()
               lvlsum += n.val
               if n.left: dq.append(n.right)
               if n.right: dq.append(n.left)
            a.append(lvlsum)
        if len(a) < k:
            return -1
        else:
            a.sort(reverse=True)
            return a[k-1]
\end{lstlisting}
\end{prompt_green}

\begin{prompt_green}
\textbf{Debugging Explanation: } Fixed iteration on children nodes and indexing for kth largest element
\end{prompt_green}

\begin{prompt_green}
\textbf{Test Results: } 
Run time: N/A\\
Run Memory: 16340000\\
Test Cases Results: \seqsplit{100000000000000000000000000000000000000000000}
\end{prompt_green}

\begin{prompt_green}
\textbf{Test Decision: } 
False
\end{prompt_green}

\vspace{10pt}
\noindent\textbf{5. GPT-3.5-Turbo, C++, unclosed string}

\begin{prompt_blue}
\textbf{Question: } You are given a string word that consists of digits and lowercase English letters.
You will replace every non-digit character with a space. For example, "a123bc34d8ef34" will become " 123  34 8  34". Notice that you are left with some integers that are separated by at least one space: "123", "34", "8", and "34".
Return the number of different integers after performing the replacement operations on word.
Two integers are considered different if their decimal representations without any leading zeros are different.
\end{prompt_blue}

\begin{prompt_blue}
\textbf{Examples: } Input: word = "a123bc34d8ef34"\\Output: 3\\Explanation: The three different integers are "123", "34", and "8". Notice that "34" is only counted once.
\end{prompt_blue}

\begin{prompt_blue}
\textbf{Oracle Solution Code: }
\begin{lstlisting}
class Solution {
public:
   int numDifferentIntegers(string word) {
        int n = word.length();
        unordered_set<string>s;
        int i=0;
        while(i<n){
            if(isdigit(word[i]) == false){
                i++;
                continue;
            }
            string temp = "";
            while(i<n && isdigit(word[i])){
                temp+=word[i];
                i++;
            }
            int j=0;
            while(j<temp.length() && temp[j] == '0') j++;
            temp = temp.substr(j);
            s.insert(temp);
        }
        return s.size();
    }
};
\end{lstlisting}
\end{prompt_blue}

\begin{prompt_red}
\textbf{Buggy Code: }
\begin{lstlisting}
class Solution {
public:
   int numDifferentIntegers(string word) {
        int n = word.length();
        unordered_set<string>s;
        int i=0;
        while(i<n){
            if(isdigit(word[i]) == false){
                i++;
                continue;
            }
            string temp = "Start of sub-string;
            while(i<n && isdigit(word[i])){
                temp+=word[i];
                i++;
            }
            int j=0;
            while(j<temp.length() && temp[j] == '0') j++;
            temp = temp.substr(j);
            s.insert(temp);
        }
        return s.size();
    }
};
\end{lstlisting}
\end{prompt_red}

\begin{prompt_red}
\textbf{Bug Explanation: } The string temp is left unclosed after "Start of sub-string". This will cause an error.
\end{prompt_red}

\begin{prompt_green}
\textbf{GPT-3.5-Turbo Debugging: }
\begin{lstlisting}
p
class Solution {
public:
   int numDifferentIntegers(string word) {
        int n = word.length();
        unordered_set<string>s;
        int i=0;
        while(i<n){
            if(isdigit(word[i]) == false){
                i++;
                continue;
            }
            string temp;
            while(i<n && isdigit(word[i])){
                temp+=word[i];
                i++;
            }
            int j=0;
            while(j<temp.length() && temp[j] == '0') j++;
            temp = temp.substr(j);
            s.insert(temp);
        }
        return s.size();
    }
};
```
<exp>
To fix the code, I added double quotes and also removed the "Start of sub-string" as it was invalid syntax
\end{lstlisting}
\end{prompt_green}

\begin{prompt_green}
\textbf{Debugging Explanation: } To fix the code, I added double quotes and also removed the "Start of sub-string" as it was invalid syntax
\end{prompt_green}

\begin{prompt_green}
\textbf{Test Results: } 
Run time: N/A\\
Run Memory: 0\\
Test Cases Results: \seqsplit{compling error}
\end{prompt_green}

\begin{prompt_green}
\textbf{Test Decision: } 
False
\end{prompt_green}

\vspace{10pt}
\noindent\textbf{6. GPT-3.5-Turbo, Java, undefined methods}

\begin{prompt_blue}
\textbf{Question: } Given an array of integers temperatures represents the daily temperatures, return an array answer such that answer[i] is the number of days you have to wait after the ith day to get a warmer temperature. If there is no future day for which this is possible, keep answer[i] == 0 instead.
\end{prompt_blue}

\begin{prompt_blue}
\textbf{Examples: } Input: temperatures = [73,74,75,71,69,72,76,73]\\Output: [1,1,4,2,1,1,0,0]
\end{prompt_blue}

\begin{prompt_blue}
\textbf{Oracle Solution Code: }
\begin{lstlisting}
class Solution {
    public int[] dailyTemperatures(int[] temperatures) {
        
        int[] arr = new int[temperatures.length]; 
        
        ArrayDeque<int[]> st = new ArrayDeque<>(); 
        
        int i=0, j=temperatures.length-1 ; 
        
        while( j >= 0){
            
            if(st.size()==0){
                arr[i++] = 0;
            }
            else  if (st.size()>0 && st.peek()[0]>temperatures[j])
            {
                arr[i++]=1;
            } 
            else if (st.size()>0 && st.peek()[0]<=temperatures[j])
            
            {  while(st.size()>0 && st.peek()[0]<=temperatures[j]){
                    st.pop();
                    
                }
            if(st.size()==0){
                arr[i++]=0;
            }
            else{
                arr[i++]=st.peek()[1]-j;
            }
            } 
        st.push(new int[]{temperatures[j],j}); 
        j--;
        }
        
      
        int[] ans = new int[temperatures.length]; int k=0;
        for(i=arr.length-1; i>=0; i--){
            ans[k++]=arr[i];
        }
        return ans;
    }
}
\end{lstlisting}
\end{prompt_blue}

\begin{prompt_red}
\textbf{Buggy Code: }
\begin{lstlisting}
class Solution {
    public int[] dailyTemperatures(int[] temperatures) {
        
        int[] arr = new int[temperatures.length]; 
        
        ArrayDeque<int[]> st = new ArrayDeque<>(); 
        
        int i=0, j=temperatures.length-1 ; 
        
        while( j >= 0){
            
            if(st.size()==0){
                arr[i++] = 0;
            }
            else  if (st.size()>0 && st.peek()[0]>temperatures[j])
            {
                arr[i++]=1;
            } 
            else if (st.size()>0 && st.peek()[0]<=temperatures[j])
            
            {  while(st.size()>0 && st.peek()[0]<=temperatures[j]){
                    st.pop();
                    
                }
            if(st.size()==0){
                arr[i++]=0;
            }
            else{
                arr[i++]=st.peek()[1]-j;
            }
            } 

        doSomething();

        st.push(new int[]{temperatures[j],j}); 
        j--;
        }
        
      
        int[] ans = new int[temperatures.length]; int k=0;
        for(i=arr.length-1; i>=0; i--){
            ans[k++]=arr[i];
        }
        return ans;
    }
}
\end{lstlisting}
\end{prompt_red}

\begin{prompt_red}
\textbf{Bug Explanation: } The method 'doSomething()' is called but it has not been defined anywhere in the class.
\end{prompt_red}

\begin{prompt_green}
\textbf{GPT-3.5-Turbo Debugging: }
\begin{lstlisting}
import java.util.ArrayDeque;

class Solution {
    public int[] dailyTemperatures(int[] temperatures) {
        
        int[] arr = new int[temperatures.length]; 
        
        ArrayDeque<int[]> st = new ArrayDeque<>(); 
        
        int i=0, j=temperatures.length-1 ; 
        
        while( j >= 0){
            
            if(st.size()==0){
                arr[i++] = 0;
            }
            else  if (st.size()>0 && st.peek()[0]>temperatures[j])
            {
                arr[i++]=1;
            } 
            else if (st.size()>0 && st.peek()[0]<=temperatures[j])
            
            {  while(st.size()>0 && st.peek()[0]<=temperatures[j]){
                    st.pop();
                    
                }
            if(st.size()==0){
                arr[i++]=0;
            }
            else{
                arr[i++]=st.peek()[1]-j;
            }
            } 

        doSomething();

        st.push(new int[]{temperatures[j],j}); 
        j--;
        }
        
      
        int[] ans = new int[temperatures.length]; int k=0;
        for(i=arr.length-1; i>=0; i--){
            ans[k++]=arr[i];
        }
        return ans;
    }

    // Added method for compilation
    public static void doSomething(){
        // Do nothing
    }
}
\end{lstlisting}
\end{prompt_green}

\begin{prompt_green}
\textbf{Debugging Explanation: } Added empty method doSomething() to fix a compile time error
\end{prompt_green}

\begin{prompt_green}
\textbf{Test Results: } 
Run time: 25 ms\\
Run Memory: 60564000\\
Test Cases Results: \seqsplit{111111111111111111111111111111111111111111111111}
\end{prompt_green}

\begin{prompt_green}
\textbf{Test Decision: } 
True
\end{prompt_green}

\vspace{10pt}
\noindent\textbf{7. GPT-3.5-Turbo, Python3, variable error}

\begin{prompt_blue}
\textbf{Question: } Given a 0-indexed integer array nums, find the leftmost middleIndex (i.e., the smallest amongst all the possible ones).
A middleIndex is an index where nums[0] + nums[1] + ... + nums[middleIndex-1] == nums[middleIndex+1] + nums[middleIndex+2] + ... + nums[nums.length-1].
If middleIndex == 0, the left side sum is considered to be 0. Similarly, if middleIndex == nums.length - 1, the right side sum is considered to be 0.
Return the leftmost middleIndex that satisfies the condition, or -1 if there is no such index.
\end{prompt_blue}

\begin{prompt_blue}
\textbf{Examples: } Input: nums = [2,3,-1,8,4]\\Output: 3\\Explanation: The sum of the numbers before index 3 is: 2 + 3 + -1 = 4\\The sum of the numbers after index 3 is: 4 = 4
\end{prompt_blue}

\begin{prompt_blue}
\textbf{Oracle Solution Code: }
\begin{lstlisting}
class Solution:
    def findMiddleIndex(self, nums: List[int]) -> int:
        def dnc(l, r, lo, hi):
            if l == r - 1:
                return l if lo == hi else -1
            
            mid = (l + r) // 2
            right = sum(nums[mid:r])
            left = sum(nums[l:mid])

            left_ind = dnc(l, mid, lo, hi + right)
            return left_ind if left_ind != -1 else dnc(mid, r, lo + left, hi)
        return dnc(0, len(nums), 0, 0)
\end{lstlisting}
\end{prompt_blue}

\begin{prompt_red}
\textbf{Buggy Code: }
\begin{lstlisting}
class Solution:
    def findMiddleIndex(self, nums: List[int]) -> int:
        def dnc(l, r, lo, hi):
            if l == r - 1:
                return l if lo == hi else -1
            
            mid = (l + r) // 2
            right = sum(nums[mid:r])
            left = sum(nums[l:mid])

            left_ind = dnc(l, mid, lo, hi + left)
            return left_ind if left_ind != -1 else dnc(mid, r, lo + left, hi)
        return dnc(0, len(nums), 0, 0)
\end{lstlisting}
\end{prompt_red}

\begin{prompt_red}
\textbf{Bug Explanation: } The bug is in the line "left\_ind = dnc\(l, mid, lo, hi + left\)" where 'left' should be replaced with 'right'. This error will result in incorrect values for the left and right sums, causing the function to incorrectly identify the middle index or terminate prematurely.
\end{prompt_red}

\begin{prompt_green}
\textbf{GPT-3.5-Turbo Debugging: }
\begin{lstlisting}
class Solution:
    def findMiddleIndex(self, nums: List[int]) -> int:
        def dnc(l, r, lo, hi):
            if l == r:
                return l if lo == hi else -1
            
            mid = (l + r) // 2
            right = sum(nums[mid:r])
            left = sum(nums[l:mid])

            left_ind = dnc(l, mid, lo, hi + left)
            return left_ind if left_ind != -1 else dnc(mid + 1, r, lo + left, hi)
        return dnc(0, len(nums), 0, 0)
\end{lstlisting}
\end{prompt_green}

\begin{prompt_green}
\textbf{Debugging Explanation: } The condition for termination was wrong, changed r-1 to r and mid to mid+1 because mid is already included.
\end{prompt_green}

\begin{prompt_green}
\textbf{Test Results: } 
Run time: N/A\\
Run Memory: 16508000\\
Test Cases Results: \seqsplit{001111111001111100111111110111100111111111100110000000000000000000000000000010001100011111101010101001000010000000000000000000000000000101010111010110011001001001110000000000000000000000000000000010000100101100000010000000000000000000010111111111111111111111111000000000000000000001111100000111}
\end{prompt_green}

\begin{prompt_green}
\textbf{Test Decision: } 
False
\end{prompt_green}

\vspace{10pt}
\noindent\textbf{8. GPT-3.5-Turbo, C++, double bugs}

\begin{prompt_blue}
\textbf{Question: } Given two strings s and goal, return true if you can swap two letters in s so the result is equal to goal, otherwise, return false.
Swapping letters is defined as taking two indices i and j (0-indexed) such that i != j and swapping the characters at s[i] and s[j].

For example, swapping at indices 0 and 2 in "abcd" results in "cbad".
\end{prompt_blue}

\begin{prompt_blue}
\textbf{Examples: } Input: s = "ab", goal = "ba"\\Output: true\\Explanation: You can swap s[0] = 'a' and s[1] = 'b' to get "ba", which is equal to goal.
\end{prompt_blue}

\begin{prompt_blue}
\textbf{Oracle Solution Code: }
\begin{lstlisting}
class Solution {
public:
    bool buddyStrings(string s, string p) {
        int n=s.length(), m=p.length();
        if(n!=m) return false;
        
        vector<int> freq(26,0);
        
        int first=-1, second=-1;
        for(int i=0;i<n;i++){
            if(s[i] != p[i]){
                if(first==-1){
                    first=i;
                }else if(second==-1){
                    second=i;
                }else{
                    return false; // more than 2 mismatching characters
                }
            }
            freq[s[i]-'a']++;
        }
        
		// similar strings
        if(first == -1 && second == -1){
            for(int i=0;i<26;i++) if(freq[i] > 1) return true; // can swap similar characters
            return false; // all characters occur only once
        }else if(second == -1){
            return false; // exactly 1 mismatch
        }
        
        return (s[first]==p[second] && s[second]==p[first]); // check similarity on 2 mismatched indexes
    }
};
\end{lstlisting}
\end{prompt_blue}

\begin{prompt_red}
\textbf{Buggy Code: }
\begin{lstlisting}
class Solution {
public:
    bool buddyStrings(string s, string p) {
        int n=s.length(), m=p.length();
        if(n!=m) return false;
        
        vector<int> freq(26,0);
        
        int first=-1, second=-1;
        for(int i=0;i<=n;i++){
            if(s[i] != p[i]){
                if(first==-1){
                    first=i;
                }else if(second==-1){
                    second=i;
                }else{
                    return false;
                }
            }
            freq[s[i]-'z']++; 
        }
        
		if(first == -1 && second == -1){
            for(int i=0;i<26;i++) if(freq[i] > 1) return true;
            return false;
        }else if(second == -1){
            return false;
        }
        
        return (s[first]==p[second] && s[second]==p[first]);
    }
};
\end{lstlisting}
\end{prompt_red}

\begin{prompt_red}
\textbf{Bug Explanation: } indexing error+ operation error
\end{prompt_red}

\begin{prompt_green}
\textbf{GPT-3.5-Turbo Debugging: }
\begin{lstlisting}
class Solution {
public:
    bool buddyStrings(string s, string p) {
        int n=s.length(), m=p.length();
        if(n!=m) return false;
        
        vector<int> freq(26,0);
        
        int first=-1, second=-1;
        for(int i=0;i<n;i++){ // modified from i<=n to i<n
            if(s[i] != p[i]){
                if(first==-1){
                    first=i;
                }else if(second==-1){
                    second=i;
                }else{
                    return false;
                }
            }
            freq[s[i]-'a']++;   // modified from 'z' to 'a'
        }
        
		if(first == -1 && second == -1){
            for(int i=0;i<26;i++) if(freq[i] > 1) return true;
            return false;
        }else if(second == -1){
            return false;
        }
        
        return (s[first]==p[second] && s[second]==p[first]);
    }
};
\end{lstlisting}
\end{prompt_green}

\begin{prompt_green}
\textbf{Debugging Explanation: } 1. Modified the loop condition from i<=n to i<n to prevent accessing out of range.
2. Changed the index calculation in freq[] from s[i]-'z' to s[i]-'a' to get the correct index of the frequency array.
\end{prompt_green}

\begin{prompt_green}
\textbf{Test Results: } 
Run time: 5 ms\\
Run Memory: 7144000\\
Test Cases Results: \seqsplit{111111111111111111111111111111111111111}
\end{prompt_green}

\begin{prompt_green}
\textbf{Test Decision: } 
True
\end{prompt_green}

\end{document}